# Extreme high lattice-misfit superalloys with regular cubic L1$_2$ particles and excellent creep resistance

Zhida Liang[1,2,*], Andreas Stark[1], Florian Pyczak[1]

1. Institute of Materials Physics, Helmholtz-Zentrum Hereon, Max-Planck-Strasse 1, Geesthacht 21502, Germany
2. Laboratory for Electron Microscopy, Karlsruhe Institute of Technology, Engesserstraße 7, Karlsruhe 76131, Germany

∗ Corresponding author: Zhida Liang, zhida.liang@outlook.com

**Abstract**

In novel Co and CoNi based superalloys, the creep resistance is limited at high temperatures due to low lattice misfit and solvus temperature. In this study, we combined the advantages of Co-Ti (high lattice misfit and solvus temperature) and Co-Al based superalloys (cuboidal precipitates) by using Ti to substitute Al in alloys of Co-30Ni-(12.5-x)Al-xTi-2.5Mo-2.5W (at.%) composition. With high Ti content, the alloys obtained extreme high lattice misfit (≥ 1.3 %) and solvus temperature (> 1150 °C). During aging at 900 °C, alloys with high Ti/Al ratio exhibited a lower γ' precipitate coarsening rate resulting from their lower γ'/γ interfacial energy and higher lattice misfit. In addition, high Ti/Al ratio brought higher γ' volume fraction and excellent mechanical properties, such as higher yield stress and better creep resistance. However, at high temperature of 1100 °C, the cubic γ' phase was decomposed into deleterious η phase with D0$_{24}$ structure if the Ti/Al ratio exceeded 1. Based on this, we outreached new alloys design with a high content of Cr and Ta and appropriate Ti/Al ratio, *i.e.*, Ti/Al ratio ≤ 1. The newly designed alloys still have high solvus temperature (> 1200 °C) and exhibit high lattice misfit (> 1.2 %) as Co-12Ti (at.%) superalloys but more regular cubic γ′ precipitates and significantly better creep resistance than superalloys Co-9Al-9W and Co-9Al-9W-2Ti at 850 and 950 °C. Nevertheless, compared with creep resistance of Ni based superalloys, our newly designed alloys still need to be further improved, especially in the 1000 and 1050 °C range.

**Introduction**

Hardening of Co-based superalloys by means of intermetallic phases of the γ′ type $Co_3Ti$ has been demonstrated in the Co-Ti, Co-Ti-Cr, Co-Ti-Mo, Co-Ti-W, Co-Ti-V and Co-Ti-Re systems [1-6]. The $L1_2$ $Co_3Ti$ hardened Co-Ti based superalloys usually have extremely high solvus temperatures. However, due to the very high lattice misfit (0.75 - 1.67 %) between γ and γ′ phases in the Co-Ti system, the morphological instability of the $L1_2$-$Co_3Ti$ phase at high temperatures formed in these systems prevented them from achieving commercial status. An example of the typical irregular rod/elongated shape of the γ′ precipitates. Zenk et al. [2] reported that the shape of γ′ precipitates becomes increasingly cuboidal with increasing Cr content from 5 at.% to 15 at.%. The main reason may be the reduction of the γ/γ′ lattice misfit from the 0.75 - 1.67% range to 0.54% with Cr additions. In addition, Mo, W, V and Re [3-6] were found to stabilize the $L1_2$-$Co_3Ti$ phase and were also able to make the shape of γ′ precipitates more cuboidal. The γ/γ′ lattice misfits in Co-12Ti, Co-12Ti-4Mo, Co-15Ti-3W and Co-12Ti-5Re were measured to be 1.17 % [3], 1.01 % [3], 0.69 % [5] and 0.50 % [6], respectively. Reduction of the γ/γ′ lattice misfit seems to be a necessary prerequisite to improve the morphological stability of γ′ precipitates. An overview of the above-mentioned influences on the microstructure of Co-Ti-based superalloys is also shown in **Fig. S1**.

In 2006, the existence of a $L1_2$ structured $Co_3(Al,W)$ phase was reported [7] and Co-based superalloys, based on the Co-Al-W system, have gained worldwide scientific attentions. In 1939, the $L1_2$ structure has been reported in Co-Al alloys by Bradley and Seager [8]. Omori *et.al.* [9] also reported that the addition of Al may promote the ordering in Co-Al alloys, evidenced by the cuboidal precipitates in a Co-14Al alloy. However, $L1_2$-$Co_3Al$ phase is not stable at high temperature. Therefore, 3rd element additions become necessary to stabilize $L1_2$-$Co_3(Al,X)$ phases. Until now, to the author's knowledge, only W, Mo and V were found to stabilize efficiently $L1_2$-$Co_3(Al,X)$ phases. Nevertheless, the $L1_2$-$Co_3(Al,W)$ and $L1_2$-$Co_3(Al,Mo)$ phases are metastable and decomposed into $D0_{19}$ and B2 phases after long term aging heat treatment [10, 11]. In addition, the γ/γ′ lattice misfit in Co-9Al-9W [12] and Co-30Ni-10Al-5Mo-2Ta-2Ti [13], measured as 0.54 % and 0.48 %, were relatively low compared to the $L1_2$ $Co_3Ti$ hardened Co-Ti based superalloys, according to data compiled in **Fig. S3**. However, Co-Al-based superalloys usually have more regular cuboidal precipitates, as shown in **Fig. S2**.

Besides, the solution heat treatment temperature window of Co-Al-based superalloys [11-14], *i.e.* $T_{window} = T_{solidus} - T_{γ′ solvus}$, is larger than that of Co-Ti-based superalloys [1-6], as is shown in **Fig. S4**. However, Co-Ti-based superalloys usually have higher γ′ solvus temperature. Therefore, the combination of Co-Al-based and Co-Ti-based superalloys should be very interesting if we can combine their individual advantages, *e.g.* high lattice misfit, high γ′ solvus temperature, appropriate solution heat treatment temperature window, regular cubic $L1_2$ particles, and excellent creep resistance.

In this study, we use Ti to substitute Al in Co-Al-Ti intermediate alloys, Co-30Ni-(12.5-x)-Al-xTi-2.5Mo-2.5W (at.%), based on the Co-Co$_3$Al-Co$_3$Ti ternary phase diagram in **Fig. 1,** to investigate the effect of Ti/Al ratio on the microstructure and mechanical properties of the alloys. We employed high energy X-ray diffraction (HEXRD) to characterize the lattice misfit of different alloys with different Ti/Al ratio. We finally found that extremely high lattice misfit alloys can be realized with regular cubic $L1_2$ particles by using a high Ti/Al ratio. In complex alloys, Cr is one important element to improve oxidation and corrosion resistance. However, in Co-Al-W based and Co-Al-Mo based superalloys, high lattice misfit usually does not exist if high contents of Cr are added due to altered W and Mo partition behavior [15, 16]. In this study, we identified alloy compositions, which do not suffer from this problem. They exhibit extremely high lattice misfit while combining a high content of Cr and Ta with a high Ti/Al ratio and achieved excellent creep resistance compared with traditional Co-Al-W based superalloys, which was never reported before.

## 2. Experiments and Methods

### 2.1 Materials

The alloys under investigation are γ′-phase hardened polycrystalline CoNi-based superalloys named 0Ti, 2Ti, 4Ti, 6Ti and 8Ti with the nominal chemical composition Co-30Ni-(12.5-x)Al-xTi-2.5Mo-2.5W (x= 0, 2, 4, 6 and 8, at.%). The alloys were melted at least 7 times under argon in the form of a 70 g ingot button on a water-cooled copper hearth using a laboratory-scale vacuum arc melting unit. The as-cast material was homogenized at 1250 ºC for 24 h, followed by ageing at 900 °C in air for 220 h, and subsequently air cooled to room temperature. In addition, we designed a series of compositionally more complex CoNi based alloys with high Ti/Al ratio and high contents of Cr, Co-35Ni-15Cr-5Al-5Ti-2Ta (at.%), Co-35Ni-9.5Cr-6.5Al-5.5Ti-1.5Ta-1.5W (at.%), Co-35Ni-9.5Cr-6Al-5.5Ti-1.5Ta-2W (at.%) and Co-35Ni-9.5Cr-6Al-5.5Ti-2Ta-2.5W (at.%), named CoNiCr-2Ta, CoNi-S1, CoNi-S2 and CoNi-S3, respectively. These new alloys also experienced the same heat treatment as above.

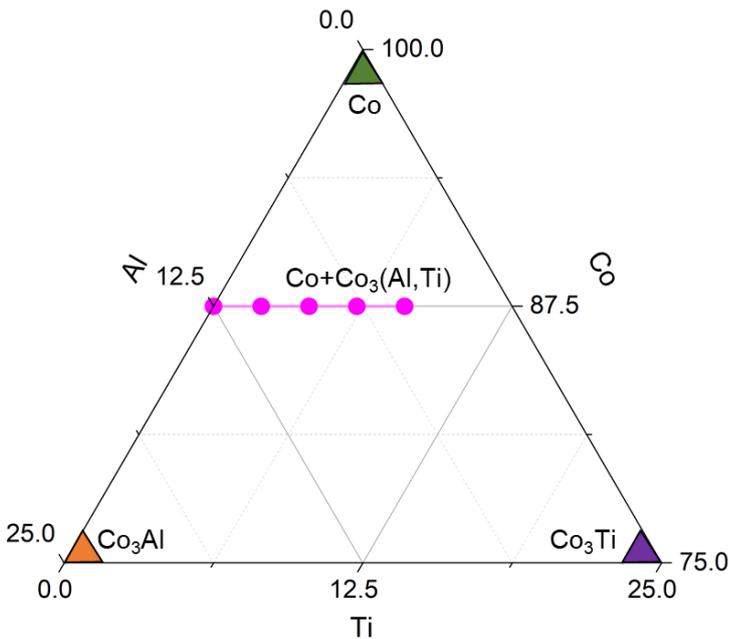

**Fig. 1.** The Co-Co$_3$Al-Co$_3$Ti ternary phase diagram (The pink dots are Al and Ti composition (at.%) of alloys in this study).

### 2.2 Mechanical testing

Compression tests to determine the 0.2% proof stress of the alloys were performed initially using a strain-controlled closed-loop machine MTS 810 (MTS System Cooperation) operated at strain rates ($\dot{\varepsilon} = 10^{-4}$ s$^{-1}$) and temperatures of 25 °C ≤ T ≤ 800 °C. Cylindrical compression specimens with a gauge length of 7.5 mm and a diameter of 5 mm were machined from the standard heat-treated material. The sample deformation was monitored by an extensometer attached over a 21 mm gauge length and the extension was used as a feed-back parameter.

Compression creep tests of the alloys were performed using a Satec Systems constant load machine with a lever arm ratio of 16:1. The creep tests in alloys 0Ti, 4Ti and 8Ti were performed in air, at 850 °C with compressive stresses of 480 MPa. The creep tests in the newly designed alloy CoNiCr-2Ta were carried out at 850 °C with compressive stresses of 350, 400 and 480 MPa and at 950 °C with compressive stresses of 150, 200 and 241 MPa in air. The alloy CoNi-S3 were carried out at 850 °C with compressive stresses of 480, 500 and 600 MPa and 950 °C with compressive stresses of 241, 300 and 350 MPa in air. The creep tests were stopped manually after a creep strain of 2.5% was reached.

**2.3 Microstructural characterization**

The γ/γ′ two-phase microstructures of the aged samples were characterized by backscattered electron (BSE) imaging in a scanning electron microscope (FE-SEM Gemini, Zeiss, Germany).

For transmission electron microscopy (TEM) sample preparation, 3 mm disks were mechanically polished to a thickness of about 75 μm. The final samples were thinned to electron transparency by a twin jet electro polishing unit using the Struers A3 electrolyte with a voltage of 32 V at -38 °C. A Thermo Fisher Scientific Talos Osiris 200 operated at 200 kV and equipped with a Super-X Energy Dispersive X-ray Spectroscopy (EDS) detector and the Esprit 2.3 software was used to investigate the compositions of the γ and γ′ phases of the aged samples. Atom force microscopy (AFM) measurements were done using a Bruker Dimension ICON atomic force microscope.

**2.4 High Energy X-ray diffraction**

The sample was measured by High Energy X-ray diffraction (HEXRD, λ = 0.124 Å) at the synchrotron beamline HEMS run by Helmholtz Zentrum Hereon at the PETRA III storage ring of the Deutsches Elektronen-Synchrotron (DESY, Hamburg, Germany). The beamline operates at an energy of 100 keV adjusted by a double crystal monochromator, which allows for measuring with complex sample environment while maintaining the possibility to access a wide **q**-range. The sample to detector distance was fixed at 1600 mm in order to obtain higher angular resolution at small diffraction angles, ideal for coherent structure characterization and crystallography. An area detector Perkin-Elmer XRD1621 was employed to record diffraction spots which is an amorphous silicon flat panel detector with CsI scintillator. In order to eliminate effects of insufficient grain statistics, the cylinder samples were oscillated angularly ± 90°. The (002) peaks were fitted by an open-source MATLAB code [17]. The lattice misfit value was calculated using the equation: $\delta = \frac{2(a_{\gamma'} - a_{\gamma})}{(a_{\gamma'} + a_{\gamma})}$, where $a_{\gamma'}$ is the lattice parameter of the γ' phase and $a_{\gamma}$ is the lattice parameter of the γ phase.

**2.5 CALculation of PHAse Diagrams (CALPHAD) simulation**

The composition of the γ and γ' phases is calculated in dependence of the Ti/Al ratio with the CALPHAD method as implemented in the Thermo-Calc 2019a software package using the TCNI8 database. Consequently, the element partition behaviour was analysed and was employed to make comparison with STEM-EDS experimental results.

**2.6 First-principles calculations**

Density functional theory (DFT) calculations as implemented in the Vienna ab initio simulation package (VASP) [18-20] were performed to predict the elastic properties of the $L1_2$-precipitate phases. To take the mixing of different atomic species on the two sublattices of the $L1_2$-structure into account, special quasi random structure (SQS) 2*2*2 supercells consisting of 32 atoms were constructed using the mcsqs routine of the ATAT software package [21]. SQS for the following compositions were constructed: $Co_3Al$, $Co_3(Al_{0.75},Ti_{0.25})$, $Co_3(Al_{0.5},Ti_{0.5})$, $Co_3(Al_{0.25},Ti_{0.75})$ and $Co_3Ti$.

**2.7 AI based segmentation method for the γ/γ' two-phase microstructure images**

The volume fraction and particle size of γ' phase are important parameters to analyse microstructure stability and mechanical properties at high temperature. In previous works [22, 23], most scientists used the well-established thresholding method, *e.g.*, Ostu's method, to segment the γ/γ' two-phase SEM-BSE image based on different contrast and brightness of the two phases. If two phases have a large difference in chemical composition, especially due to a strong partitioning tendency of heavy elements, two distinct and separate peaks can be found in the image greyscale histogram. However, if two phases exhibit very similar chemical composition, this segmentation method becomes unfeasible and unreliable. For example, in **Fig. 2**, the grayscale histogram of the SEM-BSE image has no clear two peaks which indicates that the two phases are very close to each other in brightness and contrast. In Ostu's method, we cannot get a reliable two-phase segmentation under such conditions. As an alternative, we tried an established machine learning approach called the gaussian mixture model. Here the pixel distributions of the two phases are represented as two clusters. If two clearly distinct clusters emerge, it is possible to do a separation by gaussian peak fitting. However, the method is also not suited to micrographs with overlapping greyscale ranges for the two phases.

The main reason is that the gaussian mixture model analysis still depends on the image greyscale histogram.

Deep learning currently changes many fields of science and everyday live. Recently, Meta AI introduced a new deep learning AI model, *i.e.*, segment Anything Model (SAM) [24], which claimed it can "cut out" any object, in any image, with a single click. In superalloys as investigated in the present work it preforms very well according to results in **Fig. 2**. The model SAM was run in open-source software QuPath [25] with CPU.

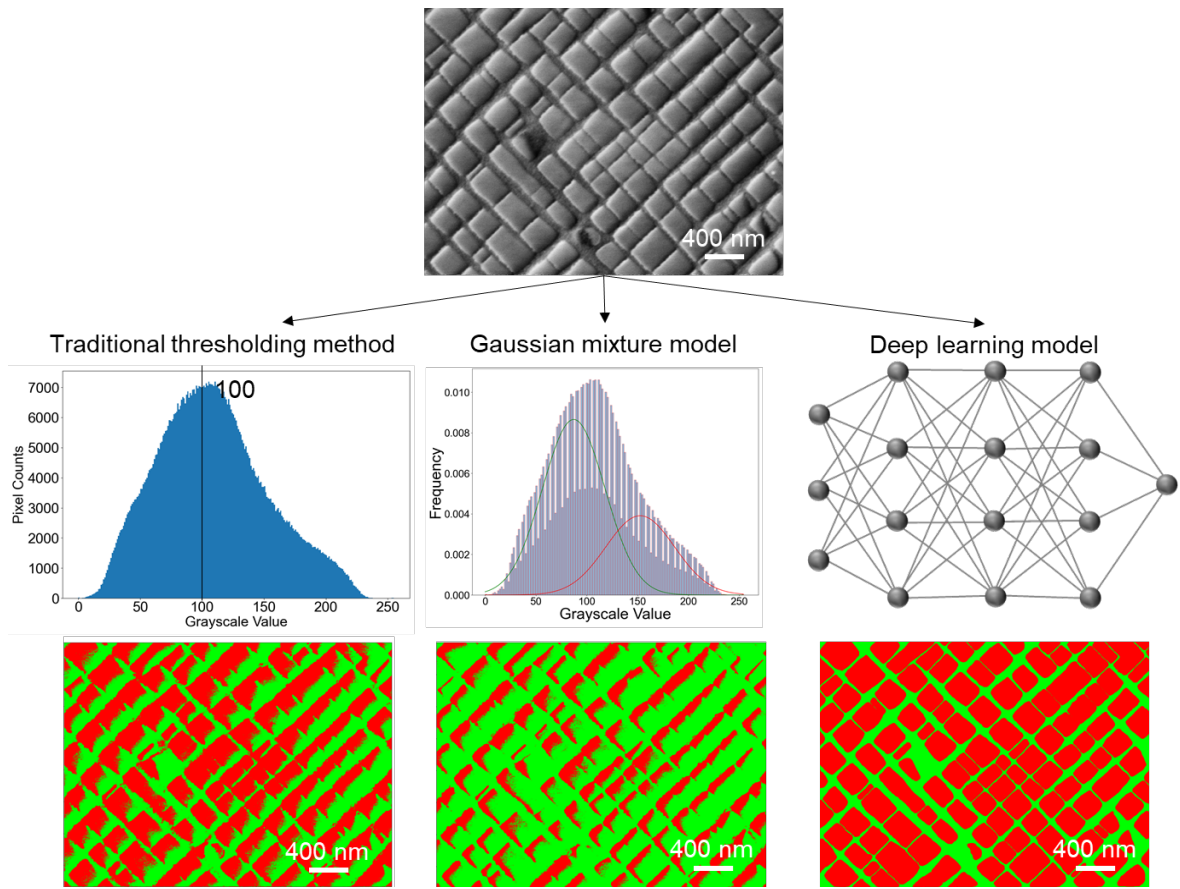

**Fig. 2.** Three segmentation methods comparison for the γ/γ′ two-phase SEM-BSE image, such as traditional thresholding method, gaussian mixture model and deep learning (Segment anything model).

### 3. Results and discussions

### 3.1. Heat treatment temperature window

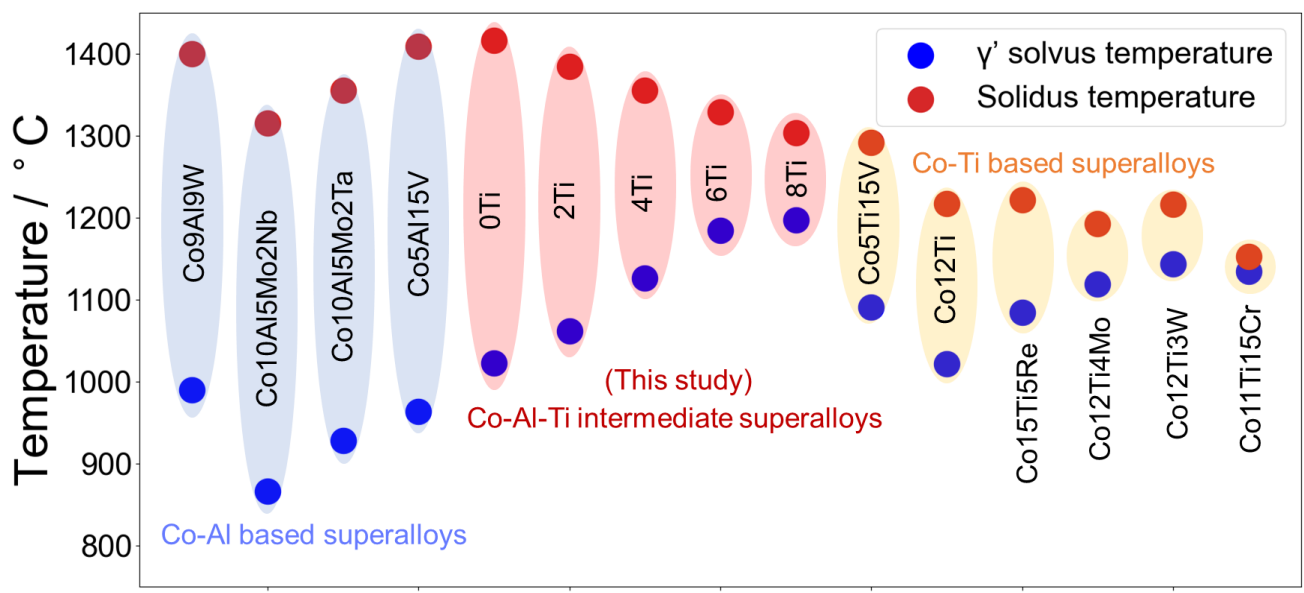

**Fig. 3.** Comparison of solution heat treatment temperature window ($T_{window} = T_{solidus} - T_{\gamma' \, solvus}$) of new Co-Al-Ti intermediate superalloys, Co-Ti-based superalloys [1-6] and Co-Al-based superalloys [11-14].

As we know, Co-Ti-based superalloys [1-6] usually have high γ′ solvus temperatures relative to their solidus temperatures so that it is not so easy to decide a suitable temperature range for solution heat treatment. Nevertheless, Co-Al-based superalloys [11-14] have lower γ′ solvus temperatures but higher solidus temperatures than Co-Ti-based superalloys. In the Co-Al-Ti alloys developed in the present work, by substituting some Ti in the alloys by Al, the solidus temperature of alloys increases and the γ' solvus temperature decreases, as is shown in **Fig. 3**. However, we can still obtain a suitable solution heat treatment temperature range combined with a high Ti concentration in those Co-Al-Ti alloys compared with Co-Ti-based superalloys. This is compatible with our ultimate goal to obtain high γ' solvus temperature together with an appropriate solution heat treatment temperature window by combining the individual advantages of Co-Ti and Co-Al based superalloys.

**3.2. Lattice parameter and lattice misfit**

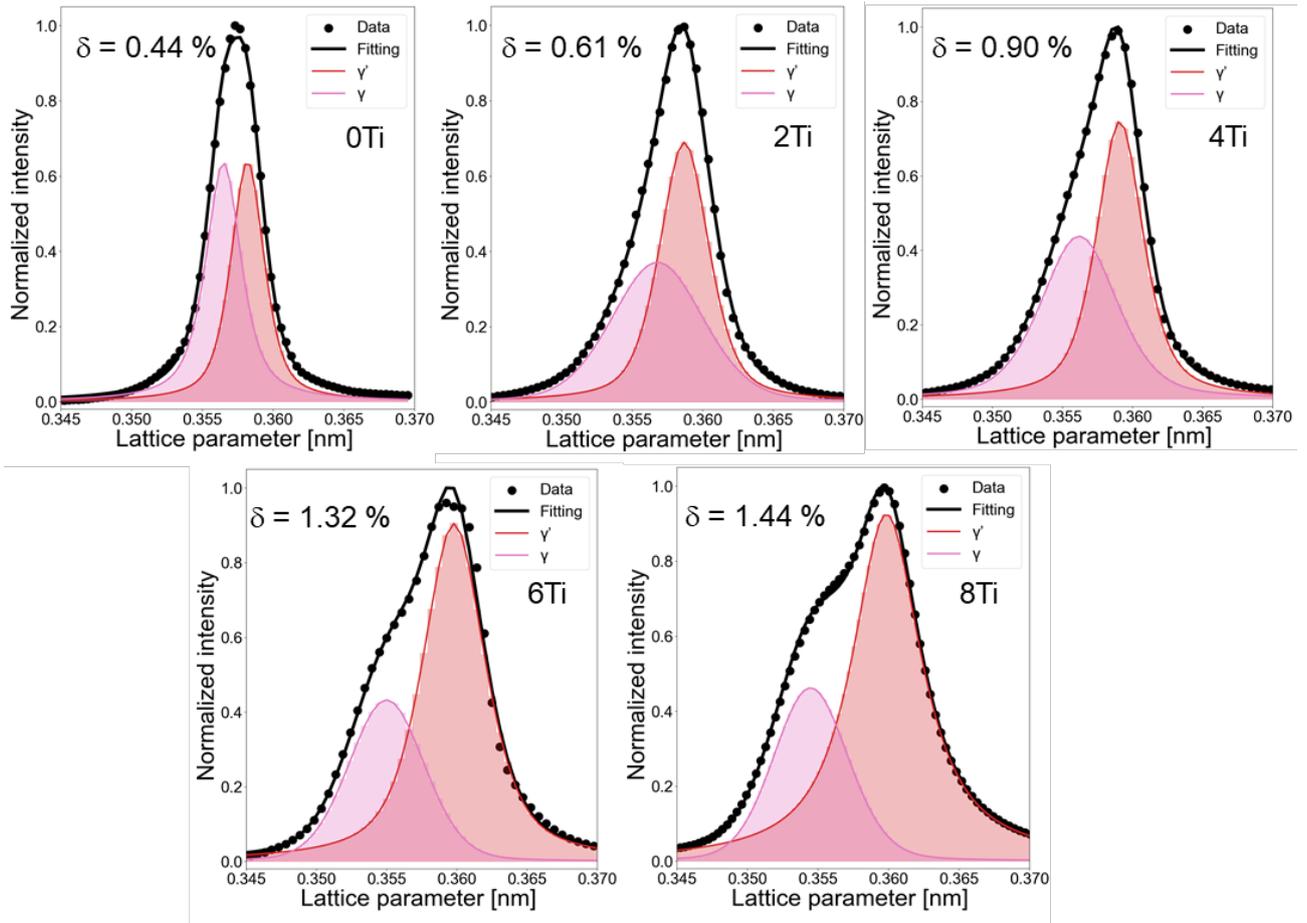

**Fig. 4.** Peaks fitting of HEXRD line profiles of (002) diffraction spots for the alloys with different Ti concentrations (0Ti, 2Ti, 4Ti, 6Ti and 8Ti) after a 220 h aging heat treatment at 900 °C.

To illustrate the change of the constrained γ/γ' lattice misfit with Ti/Al ratio in the alloys developed in the present work, HEXRD line profiles of the (002) diffraction spots measured at room temperature together with fitted peaks of the phases γ and γ′are shown in **Fig. 4**. From the fitted peaks, we obtained the average volume fraction of the γ' phases using the composite trapezoidal rule [26], *i.e.*, 47%, 53%, 53%, 66% and 69% in the alloys 0Ti, 2Ti, 4Ti, 6Ti and 8Ti, respectively, which is consistent with results from STEM-EDS analysis described below. In addition, the peak shapes become more asymmetric with Ti concentration increasing, which indicates the increasing difference between the γ and γ' phase lattice parameters. The lattice parameter and resultant lattice misfit in dependence of Ti concentration are shown in **Fig. 5(a)**. As the Ti/Al ratio increases, the lattice parameter of γ' phases rises, while the lattice parameter of γ phases exhibits a reverse trend. Consequently, the lattice misfit increases with rising Ti concentration, a trend consistent with findings in Ni based superalloys [27].

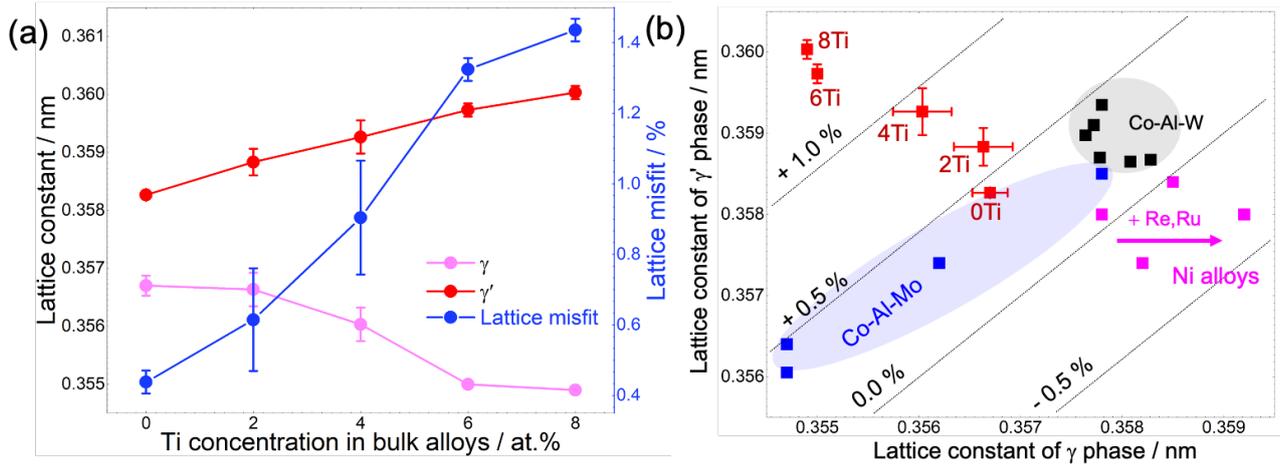

**Fig. 5.** (a) γ and γ′ lattice parameters and lattice misfit derived from (002) diffraction spots as function of Ti concentration. (b) Lattice parameter of the alloys investigated by HEXRD measurements together with literature data of Co-Al-Mo based [29], Co-Al-W based [30] and 1st-4th generation Ni-based superalloys [31] for comparison. (The diagonal lines in the diagram represent isovalue lines of γ/γ' misfit.)

**Fig. 5(b)** provided us a larger overview regarding lattice misfit comparison in different superalloy systems, such as Co-Al-W, Co-Al-Mo and traditional 1st-4th generation Ni-based superalloys. For the Ti-free variant of our alloys, *i.e.*, Co-30Ni-12.5Al-2.5Mo-2.5W, its lattice misfit (+0.44 %) is in the same range with other Co-Al-W and Co-Al-Mo based superalloys. The lattice misfit of the Co-9Al-9W superalloy was reported to be +0.53 % [12] at room temperature and increased to +0.74 % in Co-9Al-7.5W-2Ta [28]. The lattice misfit of the Co-10Al-5Mo-2Nb superalloy was +0.31 % [29] at room temperature. In this study, when we use more Ti to replace Al, the lattice misfit starts to increase and the misfit value can reach up to +1.4 % in alloy 8Ti. This misfit value is higher than that in alloy Co-12Ti (the lattice misfit is only 1.17 % [3]). From **Fig. 5(b)**, it can also be seen that the addition of elements such as Re and Ru in Ni based superalloys tends to induce a more negative lattice misfit and conversely, in CoNi based superalloys, higher additions of Ti tend to generate a more positive lattice misfit, which brings us into a desirable range to generate more regular cuboidal γ′ precipitates and the associated better material characteristics.

### 3.3. Element partition behaviour between γ and γ' phases

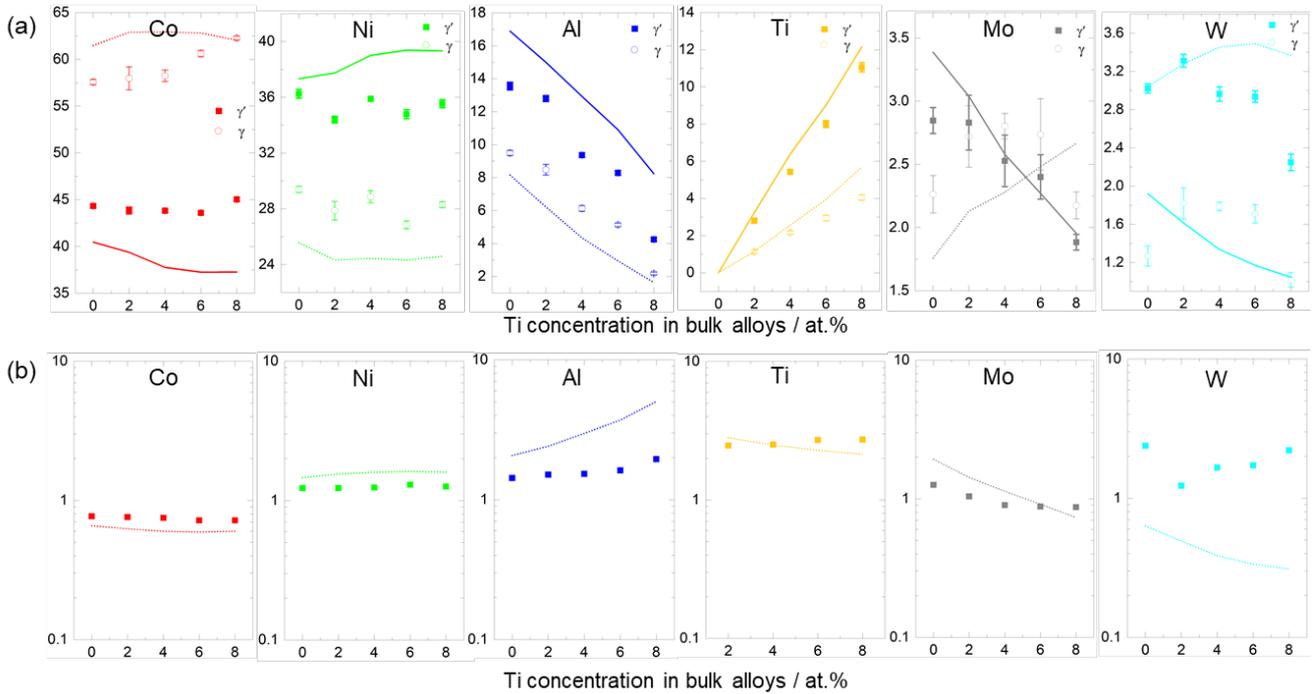

**Fig. 6**. (a) The variation of elemental composition (at.%) and (b) the element partition coefficients between the phases γ and γ′ as function of Ti concentration. (The lines represent the simulation results by CALPHAD method)

To understand element partition behaviour, the mean element concentration in the phases γ and γ′ was measured by STEM-EDS. The respective results are shown in **Fig. 6(a)** and **Table 2**. With Al substituted by Ti in series of alloys, the experimental data indicates that Co and Ni content only slightly change in the γ and γ' phases and therefore their partition coefficient stays almost constant. However, as predicted by thermodynamic modelling (lines in **Fig. 6(a)**), Ni content increases slightly in the γ' phase and decreases in the γ phase with increasing Ti content but Co has an inverse behavior. By replacing Ti with Al in alloys, of course, high Ti and low Al content existed in the γ/γ′ two phases. However, the experimental result regarding Ti partition coefficient is in contradiction with thermodynamic modelling. It is experimentally found that with Ti concentration increasing the element partition coefficient of Ti also increased but CALPHAD predicted that it would decrease. Mo concentration decreases in γ' phase but increases in γ phase with increasing Ti content, which indicates there is a transition in partitioning behavior and is Mo being a γ' phase former in Ti lean alloys becomes a slight γ phase former. W was always found to be enriched in the γ' phase in experiment, which is a completely different partition behavior compared with CALPHAD modelling which considered W as a γ matrix former.

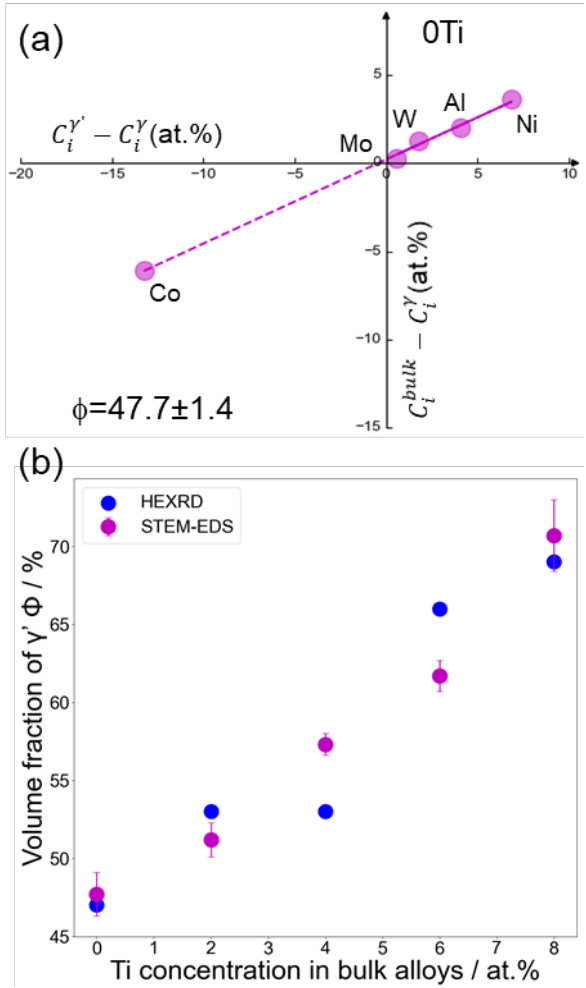

**Fig. 7.** (a) Scatter plot with linear fit between $(C_i^{\gamma'} - C_i^{\gamma})$ and $(C_i^{bulk} - C_i^{\gamma})$ for 0Ti alloy for determination of γ′ volume fraction. (b) Comparison of γ′ fraction in dependence of Ti content, as calculated from the STEM-EDS composition using a lever rule method and a HEXRD analysis approach.

The composition of the solutes determined from STEM-EDS was used to estimate the volume fraction of the γ′ and γ phases in each alloy by using the lever rule method. The expression can be written as: $C_i^{\gamma}(1 - f_{\gamma'}) + C_i^{\gamma'} f_{\gamma'} = C_i^{bulk}$, where $C_i^{\gamma}$ and $C_i^{\gamma'}$ are the compositions of solute $i$ in γ and γ', respectively, and $f_{\gamma'}$ is the volume fraction of the γ' phase. Based on this, we can derive the volume fraction of the γ' phase by: $f_{\gamma'} = (C_i^{bulk} - C_i^{\gamma})/(C_i^{\gamma'} - C_i^{\gamma})$. So, the respective values for every element were plotted and linearly fitted by the least-square method. By calculating the slope of the linear fit line, we can obtain the volume fraction of the γ' phase. The analysis results were shown in **Fig. 7(a)** and **Fig. S5**. In addition, we made a comparison of the volume fraction of the γ' phase determined based on STEM-EDS results and HEXRD results, as illustrated in **Fig. 7(b)**. STEM-EDS analysis results are in good agreement with results based on HEXRD data, which, in addition, indicates that the HEXRD peak fitting and lattice misfit analysis yield reasonable results.

**Table 2.** Phase compositions (at.%) as measured by STEM-EDS and associated partitioning coefficients for superalloys aged at 900 °C for 200 h.

| Alloys | | Co | Ni | Al | Ti | Mo | W |
|---|---|---|---|---|---|---|---|
| 0Ti | $\gamma'$ | 44.33±0.17 | 36.25±0.32 | 13.55±0.23 | – | 2.85±0.10 | 3.02±0.05 |
| | $\gamma$ | 56.08±0.45 | 28.41±0.43 | 9.50±016 | – | 2.26±0.15 | 1.27±0.11 |
| | $K_i^{\gamma'/\gamma}$ | 0.77 | 1.23 | 1.43 | – | 1.26 | 2.38 |
| 2Ti | $\gamma'$ | 43.85±0.36 | 34.40±0.24 | 12.80±0.19 | 2.81±0.13 | 2.83±0.23 | 3.31±0.07 |
| | $\gamma$ | 57.97±1.23 | 27.87±0.66 | 8.48±0.32 | 1.14±0.10 | 2.72±0.24 | 1.82±0.16 |
| | $K_i^{\gamma'/\gamma}$ | 0.76 | 1.23 | 1.51 | 2.47 | 1.04 | 1.23 |
| 4Ti | $\gamma'$ | 43.82±0.15 | 35.89±0.17 | 9.37±0.14 | 5.43±0.10 | 2.53±0.20 | 2.96±0.07 |
| | $\gamma$ | 58.24±0.61 | 28.87±0.43 | 6.13±0.18 | 2.17±0.09 | 2.80±0.10 | 1.79±0.05 |
| | $K_i^{\gamma'/\gamma}$ | 0.75 | 1.24 | 1.53 | 2.51 | 0.90 | 1.66 |
| 6Ti | $\gamma'$ | 43.59±0.17 | 34.32±1.22 | 8.29±0.15 | 8.00±0.19 | 2.40±0.18 | 2.94±0.06 |
| | $\gamma$ | 60.62±0.34 | 26.85±0.29 | 5.13±0.10 | 2.95±0.15 | 2.74±0.28 | 1.71±0.10 |
| | $K_i^{\gamma'/\gamma}$ | 0.72 | 1.30 | 1.62 | 2.71 | 0.88 | 1.72 |
| 8Ti | $\gamma'$ | 45.03±0.22 | 35.54±0.29 | 4.25±0.12 | 11.06±0.25 | 1.88±0.06 | 2.25±0.09 |
| | $\gamma$ | 62.27±0.21 | 28.31±0.21 | 2.18±0.09 | 4.05±0.14 | 2.18±0.11 | 1.02±0.08 |
| | $K_i^{\gamma'/\gamma}$ | 0.72 | 1.26 | 1.95 | 2.73 | 0.87 | 2.21 |

## 3.4. The γ' phase stability during isothermal annealing treatment

### 3.4.1. Coarsening kinetics of γ' precipitates at 900 °C - experiment and modelling

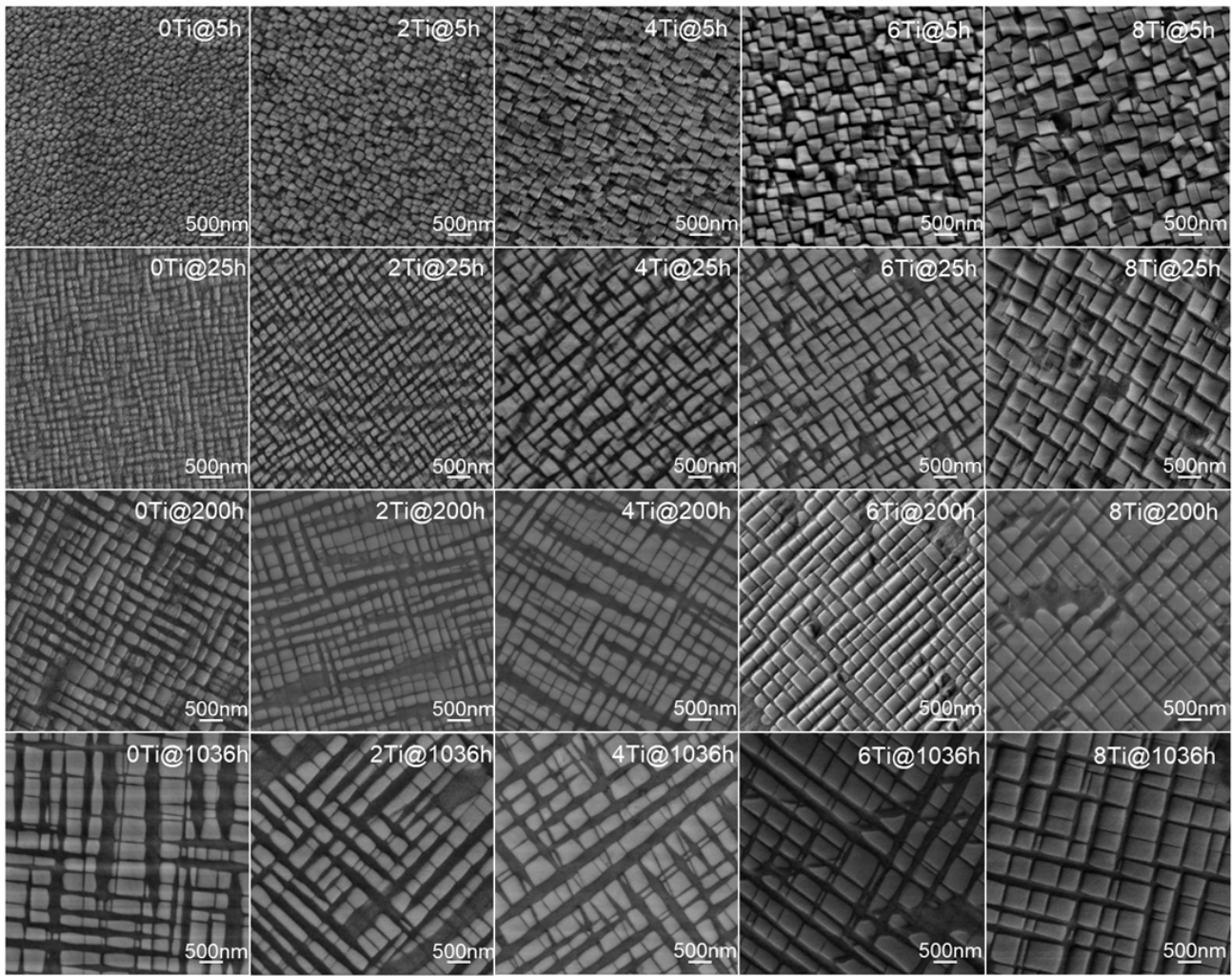

**Fig. 8.** The γ'/γ two phases microstructure of 0Ti, 2Ti, 4Ti, 6Ti and 8Ti alloys subjected to different aging heat treatments at 900 °C from 5 to 1036 h.

The coarsening rate of the γ' precipitates as employed in the theory of Ostwald ripening is one important index to characterize the γ' phase stability during isothermal annealing treatments. **Fig. 8** shows the γ'/γ two phase microstructures of 0Ti, 2Ti, 4Ti, 6Ti and 8Ti alloys subjected to different aging heat treatments at 900 °C from 5 to 1036 h. Already qualitatively investigating the BSE images in **Fig. 8**, we can draw several conclusions. Firstly, no loss of coherency seems to take place as γ' precipitates retain their cubic shape despite the high lattice misfit of all alloys, such as especially 6Ti and 8Ti, which is different to some high lattice misfit Co-Ti based [3] and Ni based superalloys [31, 32]. Secondly, there is no obvious coalescence and significant morphology changes with increasing aging time. However, with Ti/Al ratio increasing, the shape of the γ' particles becomes increasingly more cubic which should be related to the increasing lattice misfit. The third observation is that the γ' particles in the high Ti/Al ratio alloy 8Ti are bigger than those in low Ti/Al ratio alloys, *e.g.*, 0Ti, before 200 h aging time but this gap seems to become smaller at 1036 h. In order to prove this finding statistically, we segmented every γ' particle from the BSE images and obtained particles' size distribution by using a deep learning model, *i.e.*, segment anything model (SAM). The particles' size distribution and the gaussian fit to it for alloys 0Ti, 2Ti, 4Ti, 6Ti and 8Ti during aging at 900 °C for 0, 5, 25, 200, 400 and 1036 h were shown in **Fig. S6**.

To clearly see the differences of particles' size distribution of alloys 0Ti, 2Ti, 4Ti, 6Ti and 8Ti after different aging time, we use a gaussian fit to the profiles to allow comparison, which is shown in **Fig. 9(a)**. The gaussian fit showed that the coarsening rate becomes smaller with increasing Ti concentration. The coarsening of the γ' particles was considered as the rate-limiting step in superalloys [33]. Typically, such a time-dependent coarsening process is described according to the LSW model by the following power law expression:

$$\bar{r}_t^3 - \bar{r}_0^3 = kt \tag{1}$$

where $\bar{r}_t$ is the mean particle size at a time *t*, $\bar{r}_0$ is the mean particle size at the onset of coarsening (at *t* = 0), and *k* is described as the coarsening constant that is largely dependent on factors such as the volume fraction and size distribution of precipitates. Experimentally, the value *k* is obtained from the slope of the plot $\bar{r}_t^3 - \bar{r}_0^3$ versus *t*. The coarsening rate constants determined by this method for alloys 0Ti, 2Ti, 4Ti, 6Ti and 8Ti at 900 °C are shown in **Fig. 9(c)** and indicate that the coarsening rate constant *k* decreases with increasing Ti additions to the alloys, i.e., the γ' particles in high Ti/Al ratio alloys exhibit low coarsening rates.

Assuming (based on STEM-EDS observations) that the diffusion of W in the γ matrix is the rate-limiting step for precipitate coarsening, the γ/γ' interfacial energy, *σ*, can be determined as follows [34]:

$$\sigma = \frac{9kRT(C_W^{\gamma'} - C_W)^2}{8DC_W(1 - C_W)V_m} \tag{2}$$

where *D* is the diffusion coefficient of W in the matrix (7.28 × 10$^{-17}$·m$^2$·s$^{-1}$ at 900 °C [34]), $C_W$ is the equilibrium solubility (in atomic fraction) of W in the γ matrix and, $C_W^{\gamma'}$ the γ' precipitate, respectively, $V_m$ is the molar volume of the precipitate, *R* is the universal gas constant (8.31 J·mol$^{-1}$·K$^{-1}$) and *T* is the absolute temperature. The molar volume of precipitates in fcc alloys is calculated by:

$$V_m = \frac{N_A a_{\gamma'}^3}{4} \quad (3)$$

where $N_A$ is Avogadro constant (6.02214076×10²³) and $a_{\gamma'}$ is lattice constant of the γ' phases measured by HEXRD. The γ/γ' interfacial energies in the alloys as calculated by **equation (2)** are shown in **Fig. 9(d)**. The results show that the high Ti/Al ratio alloys have lower γ/γ' interfacial energy. According to literature [35], a greater γ/γ' interfacial energy $\sigma$ is usually considered to increase the driving force for coarsening. The coarsening of larger precipitates at the expense of smaller leads to a reduction of the total interfacial area which results in a reduction of total γ/γ' interfacial energy.

In addition, the coherent strain energy significantly influences the morphology transition of the γ' precipitates. Transition from sphere to cubic to more cubic to rod-like shapes and alignment along the elastically soft directions of the crystal, usually ⟨100⟩ crystal directions, takes place as they increase in size. Based on this, the equilibrium shape depends on the dimensionless parameter **L** [36]:

$$L = \frac{\delta^2 \bar{r} C_{44}}{\sigma} \quad (4)$$

where $\delta$ is the misfit between γ and γ' phases, $\bar{r}$ is the mean radius of a particle and, $C_{44}$ is a single crystal elastic constant. Based on equation (4), this dimensionless parameter in alloys 0Ti, 2Ti, 4Ti, 6Ti and 8Ti is 0.94, 26.07, 82.26, 166.63 and 216.04, respectively. Therefore, for a high content Ti, the γ' precipitates are more cubic but no rod-like shapes are observed.

On the other hand, as the precipitates grew up, the effect of the strain energy became more significant [37]. Due to the increasing coherency strains, the growth rate of the eight ⟨111⟩ branches of γ' precipitates is enhanced to form six {100} crystal planes because ⟨100⟩ were the softest directions in the matrix. The new {100} interfaces between γ and γ′ lead to decreased coherency stresses and accordingly a decreased elastic strain energy associated with the precipitate size [38]. Therefore, high lattice misfit alloys with coherent γ/γ' interfaces usually exhibit an accelerated the splitting of the γ' precipitates. By combining the effect of a high strain energy to generate cuboidal precipitates with a low interfacial energy between γ and γ' to restrict precipitate coarsening, the high Ti/Al ratio alloys can exhibit a nearly perfect cubic precipitate morphology without being prone to early adopt a rod-like precipitate shape.

However, the discussion above is in contradiction to the early stages of the aging heat treatment process. From BSE images in **Fig. 8**, at the low aging time, such as 5 and 25 h, it is obvious that the high Ti/Al ratio alloys initially have higher coarsening rates according to their larger γ' particle sizes.

We assume that this is caused by a relatively smaller lattice misfit at the begin of the aging treatment than that at the high aging time. This concept of lattice misfit changing with aging time is also reported by Collins et al. [39]. Therefore, the effect of the strain energy is very small and precipitates splitting is not obvious.

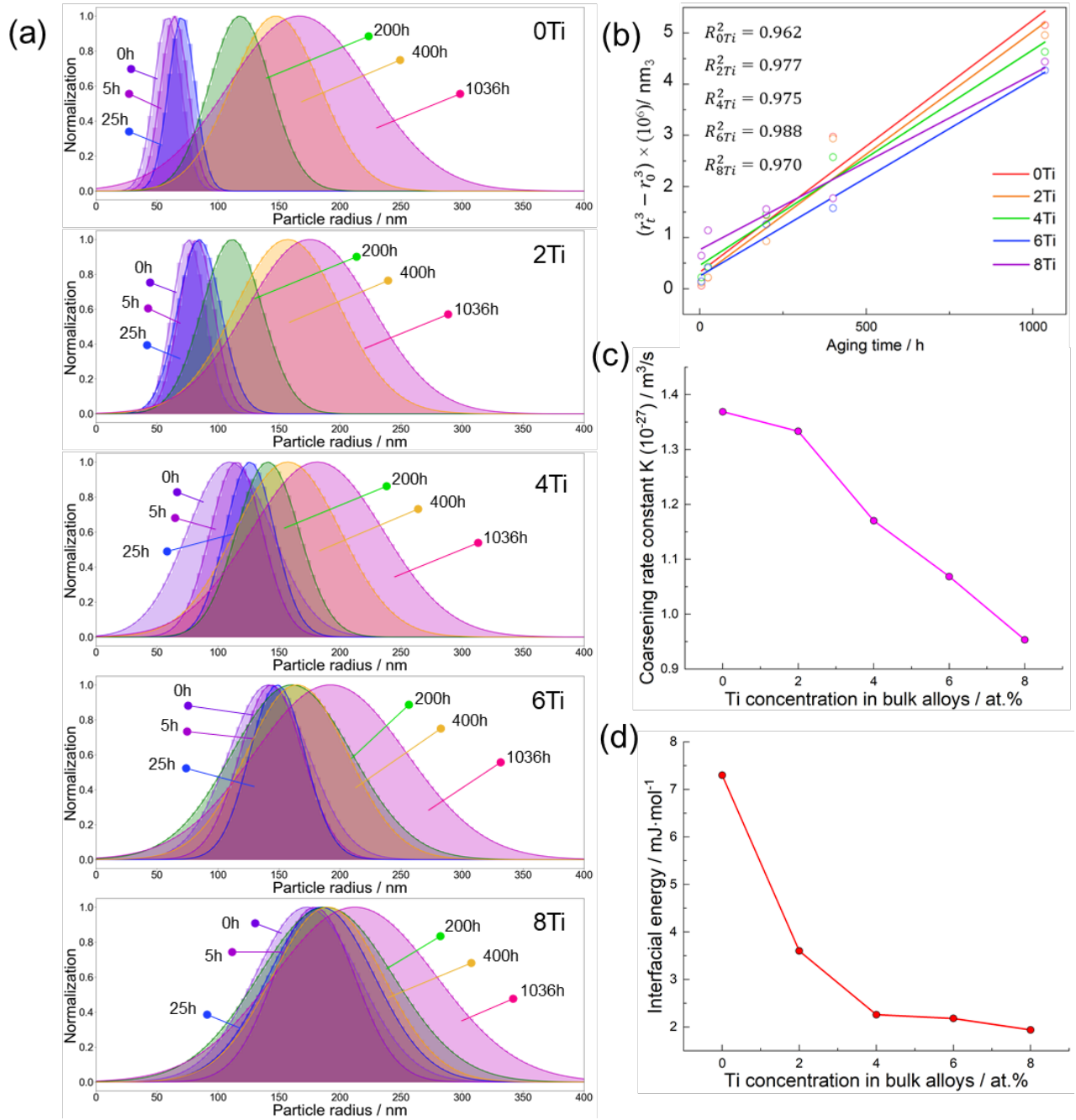

**Fig. 9.** (a) The gaussian distribution of γ′ particle sizes in alloys 0Ti, 2Ti, 4Ti, 4Ti and 8Ti. (b) Plot showing linear fit of particles size ($r^3$ in nm$^3$) versus ageing time (*t*) during isothermal annealing at 900 °C along coefficients for the quality of the linear fit. (c) LSW coarsening rate constant plotted over Ti concentration in the alloys. (d) Interfacial energy versus Ti concentration in the alloys.

### 3.4.2. The γ′ phase thermal stability at temperatures above 900 °C

To test the temperature capability of the alloys, we did annealing treatment for alloys 0Ti, 2Ti, 4Ti, 6Ti and 8Ti at the temperatures, *e.g.* of 1000 °C, and 1100 °C. According to BSE and EBSD images in **Fig. 10** (microstructures of alloys aged at 900 °C under similar conditions are shown for comparison), if the Ti concentration is lower than 8 at.%, *i.e.*, Ti:Al ≤ 1, the γ′ phase is stable at all temperatures. In the alloy 8Ti, a secondary phase η (D024, P6$_3$/mmc (194)) formed along grain boundaries at 900 °C and 1000 °C. At 1100 °C, all of the cubic γ′ phase is decomposed into η precipitates with needle morphology. Therefore, in order to obtain a stable γ′ phase, the Ti/Al ratio should be smaller than or equal to 1.

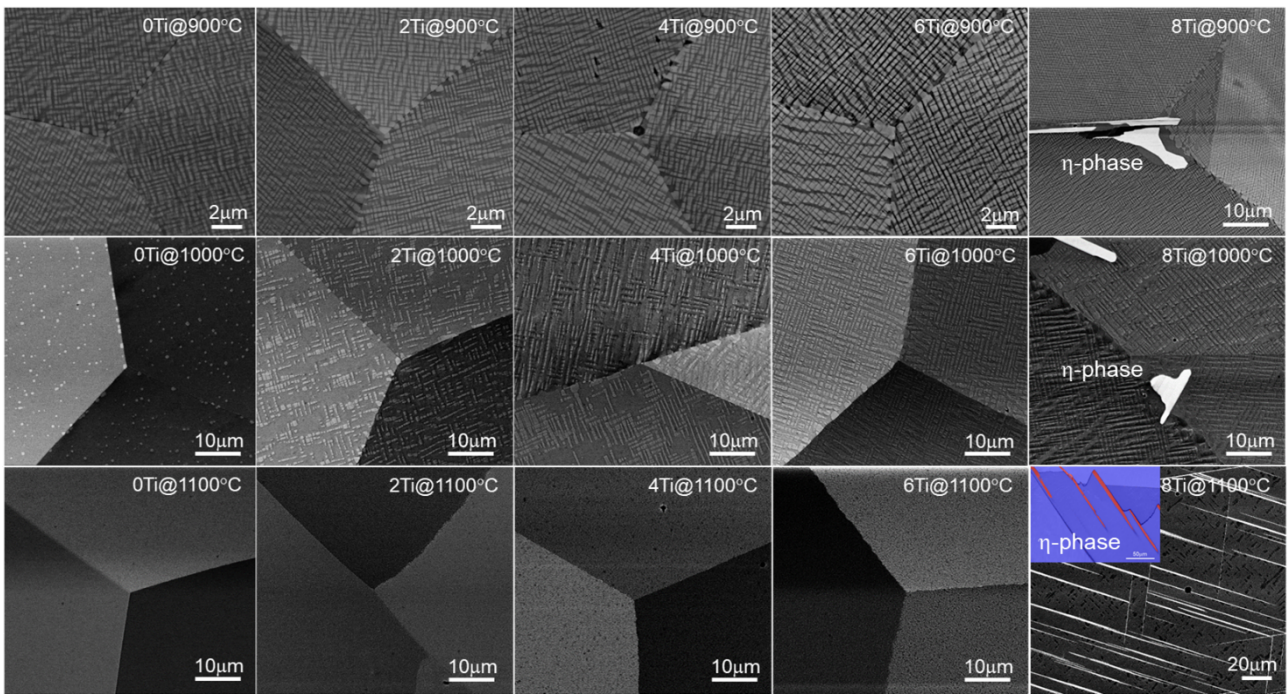

**Fig. 10**. γ′ phase morphology of specimens aged at temperatures of 900 °C, 1000 °C and 1100 °C for 200 h and pictured by SEM-BSE.

### 3.5. Mechanical properties of alloys with different Ti/Al ratio

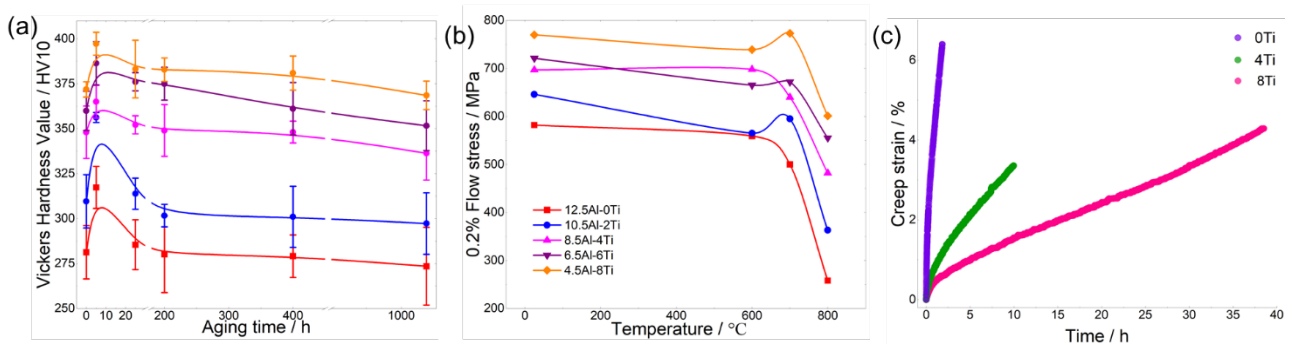

**Fig. 11.** (a) Vickers hardness measurement results (measured at 20 °C) as a function of aging time (0, 5, 25, 200, 400 and 1036 h). (b) Temperature-dependent yield strength in alloys 0Ti, 2Ti, 4Ti, 4Ti and 8Ti with 200 h aging treatment at 900 °C at different temperatures. (c) Creep test in alloys 0Ti, 4Ti and 8Ti at 850 °C with applied stress of 480 MPa.

**Fig. 11(a)** shows the evolution of Vickers hardness (HV10 measured at room temperature) for the alloys 0Ti, 2Ti, 4Ti, 6Ti and 8Ti aged for 0, 5, 25, 200, 500 and 1000 h at 850 °C. All alloys show a qualitatively similar behavior. The hardness increases mildly at the beginning of aging heat treatment and reaches the highest value at 5 h aging time. After 5 h, the hardness value starts to decrease. This phenomenon matches very well with the weak and strong pair coupling strengthening mechanism. In the weak pair case, the critical resolved shear stress can be expressed by [40]:

$$\tau_p^{Weak} = \frac{\Delta E_{APB}}{2b}\left(\sqrt{\frac{6\Delta E_{APB}\bar{r}V_{\gamma'}}{2\pi T_{ten}}} - V_{\gamma'}\right) \quad (5)$$

In the strong pair case, the critical resolved shear stress can be expressed by [40]:

$$\tau_p^{Strong} = \sqrt{\frac{3}{2}}\left(\frac{Gb}{\pi^{3/2}}\right)\frac{V_{\gamma'}^{1/2}}{\bar{r}}\sqrt{\frac{2\pi\Delta E_{APB}\bar{r}}{Gb^2} - 1} \quad (6)$$

where $\Delta E_{APB}$ is the antiphase boundary energy (kJ/mol), $\bar{r}$ is the mean particle radius (nm), $V_{\gamma'}$ is the volume fraction of γ′ phase, $G$ is the shear modulus of γ′ phase (GPa), $T_{ten} = 0.36Gb^2$ is a modified version of the dislocation line tension when the dislocation is of a mixed 60° type, $b$ = 0.248 nm is the magnitude of the Burgers vector, $L$ is the mean particle spacing. The Hill shear modulus (G) as function of Ti concentration in the γ′ phase calculated by DFT method is shown in **Table. 3**.

According to Crudden's model [41] for the calculation of antiphase boundary energy ($\Delta E_{APB}$) based on DFT calculations for multicomponent superalloys, the high Ti/Al ratio in the γ′ phase can contribute to a high $\Delta E_{APB}$ [42, 43]. Therefore, additions of high contents of Ti are beneficial for mechanical properties of alloys. The change in $\Delta E_{APB}$ was determined by linear superposition of the effect of the individual elements according to the equation:

$$\Delta E_{APB} = E_{APB}^0 + \sum_i^n (k_i x_i) \quad (7)$$

where $x_i$ is the concentration in at. % of the solute $i$ in the γ′ phase, $k_i$ is the correlation coefficient for change in $\Delta E_{APB}$, $n$ is the number of solute elements in the γ′ phase and $\Delta E_{APB}^0$ is the $\Delta E_{APB}$ for L1$_2$ Co$_3$(Al,W) (293 mJ/m$^2$m) calculated using DFT [44]. In order to know the correlation coefficient for change in $\Delta E_{APB}$ against Ti concentration in the γ′ phase, we need to know $\Delta E_{APB}$ of L1$_2$ Co$_3$(Ti,W) which is 343 mJ·m$^{-2}$ [45]. Based on this, we can infer that the correlation coefficient $k_{Ti}$ in the γ′ phase is 4 mJ m$^{-2}$/at. %. In order to simplify the calculation of the $\Delta E_{APB}$ of the γ' phase in this study, we excluded other elements' effect, *i.e.*, Co, Ni, Mo and W, because their composition changes not significantly in the γ' phase with varying Ti/Al ratio in the alloys. Therefore, the $\Delta E_{APB}$ of the γ' phase in alloys 0Ti, 2Ti, 4Ti, 6Ti and 8Ti can be estimated to be roughly 293, 304, 314, 325, and 337 mJ·m$^{-2}$/at. %, respectively.

In order to reveal the particle size strengthening of the alloys, we can plot the relationship between the critical resolved shear stress ($\tau$) and particle size, as shown in **Fig. 12**. Based on the calculations, the critical resolved shear stress increases first to reach its maximum and decreases with γ' particles further growing which matches very well with the Vickers hardness measurement results in **Fig. 11(a)**. The temperature-dependent yield strengths in alloys 0Ti, 2Ti, 4Ti, 4Ti and 8Ti are shown in **Fig. 11(b)**. Before compression experiments, the tested samples experienced a 200 h aging heat treatment at 900 °C. Their mean particle size is bigger than 100 nm based on particles size distribution in **Fig. 9(a)**. Therefore, the precipitate strengthening mechanism is strong pair-coupled strengthening in the compression test samples. Based on strong pair-coupled strengthening as plotted in **Fig. 12**, the critical resolved shear stress is higher in high Ti content alloys. The main reason is that a high content Ti will result in higher APB energy of γ' precipitates. In addition, below 800 °C, plastic deformation takes place via athermal yielding and the yield strength is mainly determined by the APB involved in the shearing of γ' particles by coupled dislocation pairs [46]. Therefore, high Ti content alloys will have higher yield stress, which is compatible with results in **Fig. 11(b)**.

**Fig. 11(c)** presents creep tests in alloys 0Ti, 4Ti and 8Ti at 850 °C with an applied stress of 480 MPa. The alloy 8Ti exhibits much higher creep resistance at 850 ºC than the low-Ti 4Ti and Ti-free 0Ti. The creep process is diffusion rate-controlled, and it is noteworthy that the alloys 0Ti, 4Ti and 8Ti contain the same amount of slow diffusing elements Mo and W. Nevertheless, high Ti/Al ratio can bring the alloys higher volume fraction of the γ' phase, higher γ'/γ lattice misfit value and higher APB energy strengthened γ' precipitates. H. Harada et.al. [47, 48] concluded that the creep resistance is increased with increasing volume fraction of γ' precipitates when this value is lower than 65 %. In addition, they also proposed that a higher γ'/γ lattice misfit will promote the formation of denser dislocation networks, which will also enhance creep resistance [49, 50]. These factors contribute to the improved creep resistance of high Ti/Al ratio alloys.

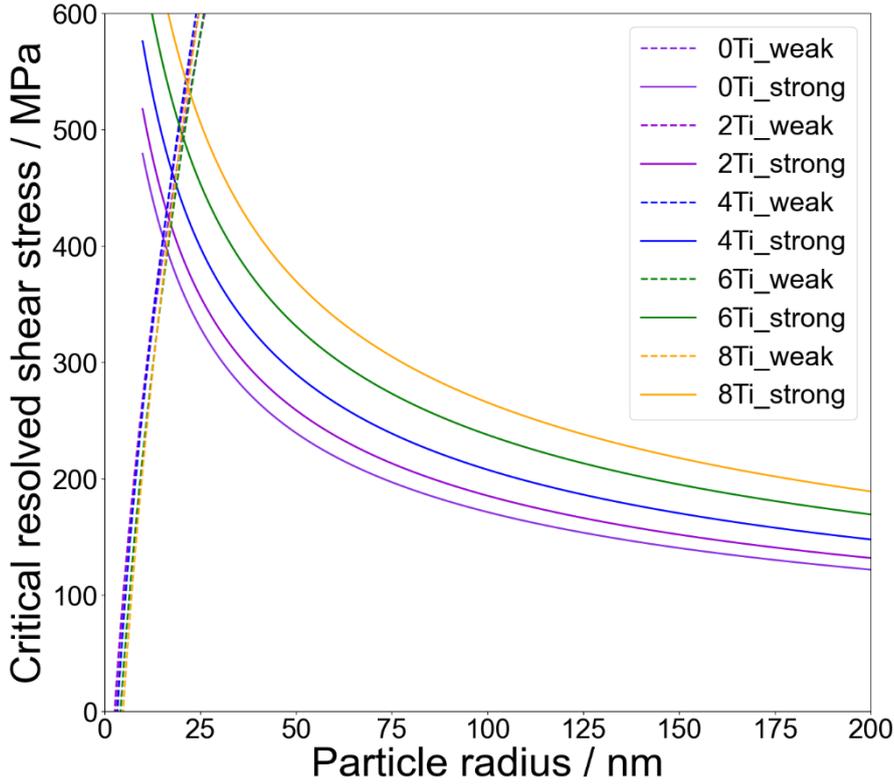

**Fig. 12.** Calculated critical resolved shear stress by weak pair-coupling and strong pair-coupling.

**Table. 3.** Calculated elastic constants ($C_{11}$, $C_{12}$, $C_{44}$) and Hill shear moduli (G) by density function theory (DFT) method.

|          | $Co_3Al$ | $Co_3Al_{0.75}Ti_{0.25}$ | $Co_3Al_{0.5}Ti_{0.5}$ | $Co_3Al_{0.25}Ti_{0.75}$ | $Co_3Ti$ |
|----------|----------|--------------------------|------------------------|--------------------------|----------|
| $C_{11}$ | 199.6    | 209.2                    | 231.1                  | 245.0                    | 275.7    |
| $C_{12}$ | 145.7    | 152.3                    | 167.1                  | 167.5                    | 171.3    |
| $C_{44}$ | 132.4    | 138.7                    | 147.8                  | 172.2                    | 155.0    |
| G        | 70.9     | 74.5                     | 80.9                   | 95.6                     | 100.3    |

### 3.6. To outreach alloy design by using high Ti/Al ratio

According to the investigations above, we can summarize that a high Ti/Al ratio in alloys causes higher lattice misfit, low γ' particle coarsening rate, higher yield strength and better creep resistance. However, if the Ti/Al ratio >1, *e.g.*, 8Ti alloy, the alloys become thermally instable at high temperatures, such as 1100 °C. The γ' phase with $L1_2$ structure will decompose into η phase with D024 structure, which is detrimental for the mechanical stability of structural components, especially in extreme environments. Therefore, in our further alloy design, we limited Ti/Al ratio ≤ 1 to guarantee better γ′ phase stability. In addition, we also added high contents Cr to improve the high temperature oxidation resistance.

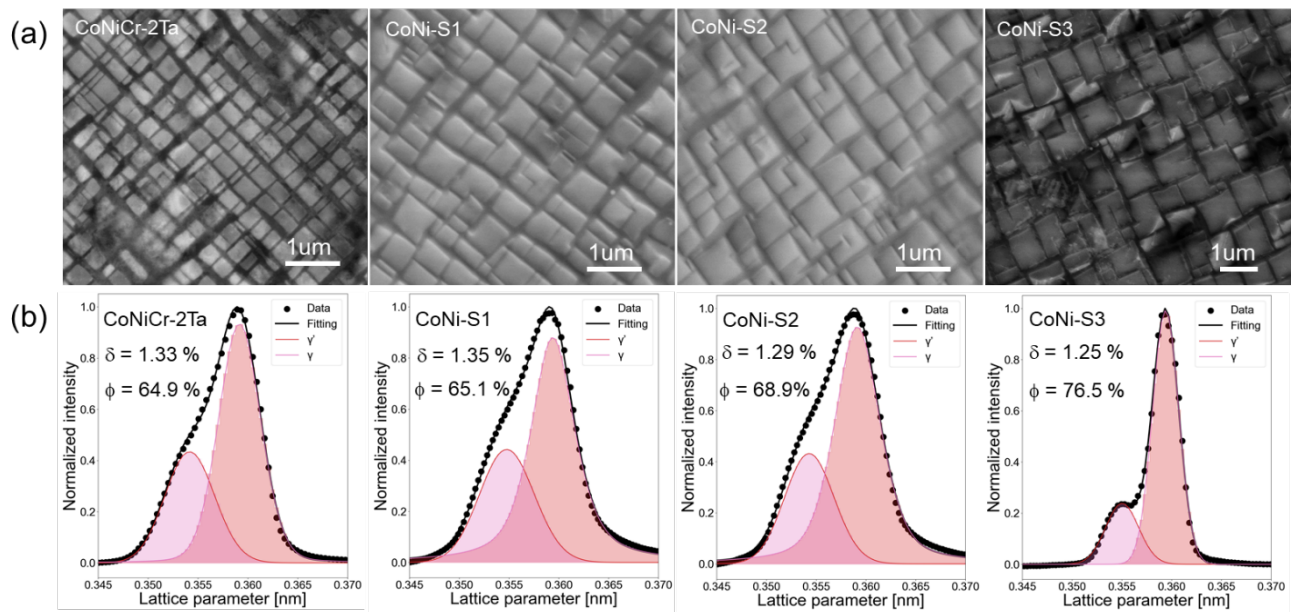

**Fig. 13.** Alloys CoNiCr-2Ta (Co-35Ni-15Cr-5Al-5Ti-2Ta, at.%), CoNi-S1 (Co-35Ni-9.5Cr-6.5Al-5.5Ti-1.5Ta-1.5W, at.%), CoNi-S2 (Co-35Ni-9.5Cr-6Al-5.5Ti-1.5Ta-2W, at.%) and CoNi-S3 (Co-35Ni-9.5Cr-6Al-5.5Ti-2Ta-2.5W, at.%) after 200 h aging heat treatment at 900 °C: **(a)** BSE images of the γ′/γ dual phase microstructure and **(b)** (002) HEXRD peaks with fitted peaks for phases γ and γ′. The lattice misfit and γ′ volume fraction derived from these fits is given in each diagram.

**Fig. 13 (a)** shows the microstructures of four optimized new alloys we designed by employing high Ti/Al ratio (close to 1) and high contents of Cr (9.5 at.%) and Ta (≥ 1.5 at.%). **Fig. 13 (b)** shows HEXRD line profiles of those new alloys of the (002) diffraction spots with fits of the γ and γ′ subpeaks. By peak fitting, we confirmed that the new alloys still have high lattice misfit and the SEM micrographs show that they exhibit regular cubic γ′ particles. This is noteworthy because the new alloy compositions also contain a high Cr content. Usually, in some Co and CoNi based superalloys [28, 29], *e.g.*, Co-Al-Mo and Co-Al-W systems, high contents of Cr will lead to a decrease of lattice misfit. The main reason why this takes not place in the alloys investigated in this work may be that we did not add γ matrix former, such as Mo. Cr additions usually influence the Mo partition ratio to tend more to the γ matrix and accordingly decrease the lattice misfit. In addition, the alloy CoNiCr-2Ta has 1.4 % lattice misfit, which is higher than Co-12Ti alloys [3] (δ = 1.17 %). However, we already know that the shape of γ′ precipitates in Co-12Ti alloys is irregular.

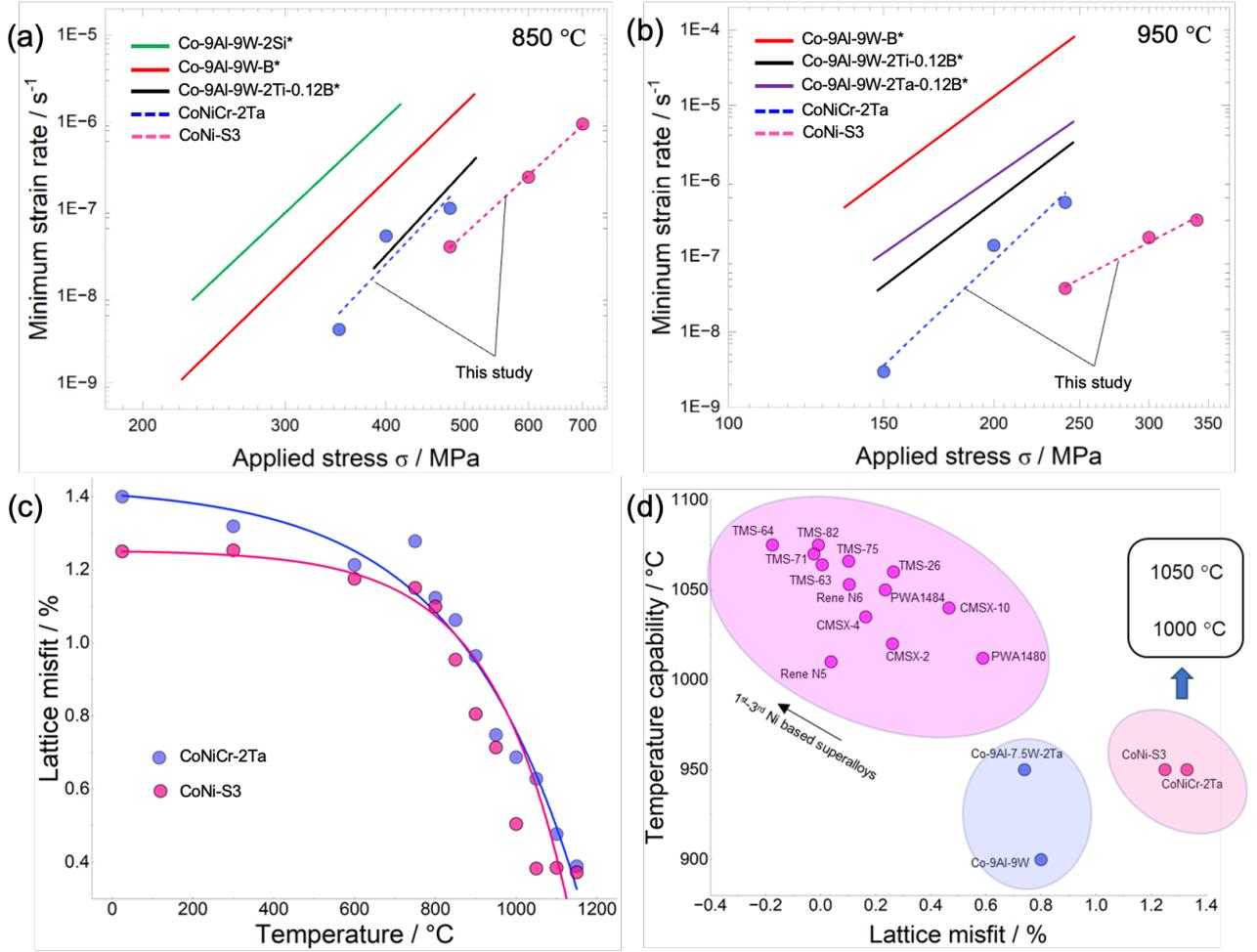

**Fig. 14.** (a) and (b) Comparison of creep strength of newly designed alloys CoNiCr-2Ta, CoNi-S3 and typical Co-9Al-9W based superalloys tested at 850 °C and 950 °C at different stresses. (c) Lattice misfit δ as a function of temperature as measured by HEXRD. (d) Comparison of new CoNiCr-based superalloys and 1st-3rd generation Ni based superalloys in terms of temperature capability [41] and lattice misfit. (Temperature capability indicates the alloys have super long creep life.)

**Fig. 14(a)** and **(b)** show the minimum creep rate $\dot{\varepsilon}$ as a function of compressive stress $\sigma$ for our CoNiCr-2Ta and CoNi-S3 alloys tested at 850 °C and 950 °C. An apparent stress exponent, $n$ = 9.7 for CoNiCr-2Ta and $n$ = 8.5 for CoNi-S3 at 850 °C and $n$ = 11 for CoNiCr-2Ta and $n$ = 6 for CoNi-S3 at 950 °C, is determined based on the power-law creep equation:

$$\dot{\varepsilon} = A \cdot \sigma^n \cdot \left(-\frac{Q}{RT}\right) \tag{8}$$

with constant *A*, applied stress in compression **σ**, stress exponent **n**, activation energy **Q**, gas constant **R** and temperature **T** in Kelvin. Therefore, CoNi-S3 has better creep strength especially at higher stress levels (lower minimum strain rate) compared with CoNiCr-2Ta. The main reason is that CoNi-S3 has a higher volume fraction of γ' phase and a small amount of W added, which has a lower diffusion rate than other elements (Ni, Cr, Al, Ti, Ta) in fcc Co [51]. Creep at high temperature is predominantly diffusion rate-controlled, although lattice misfit has also a significant effect on creep resistance [49].

The newly developed alloys perform superior compared to polycrystalline Co-base superalloys [52], Co9Al9W-B, Co9Al9W2Ti-0.12B and Co9Al9W2Ti-0.12B. In addition, these Co-base superalloys exhibit a mass density in the range of 9.4-10.2 g/cm$^3$ [52], which are apparently higher than our new alloys ($\rho_{CoNiCr-2Ta}$ = 8.24 g/cm$^3$, $\rho_{CoNi-S1}$ = 8.47 g/cm$^3$, $\rho_{CoNi-S2}$ = 8.60 g/cm$^3$ and $\rho_{CoNi-S3}$ = 8.73 g/cm$^3$). These results show clearly the potential of the new alloys since they are more creep resistant than these typical Co-base counterparts and this superiority will even be increased if specific strength would be compared due to their lower mass density. To determine the lattice misfit of the new alloys at elevated temperatures, we performed in situ heating experiment and measured lattice parameters by HEXRD. The results are shown in **Fig. 14(c)**. The γ/γ' lattice misfit **δ,** in alloy CoNiCr-2Ta, was + 1.4% at room temperature and decreased to approximately +0.68 % at 1000 °C and +0.48 % at 1100 °C. The alloy CoNi-S3 has a slightly lower lattice misfit, +0.50 % at 1000 °C and +0.38 % at 1100 °C, but CoNi-S3 has higher solvus temperature (1126 °C in CoNiCr-2Ta and 1230 °C in CoNi-S3). Above 800 °C, the lattice misfit depends mainly on the change of chemical composition particularly of the γ phase due to the proceeding dissolution of the γ' phase. In Ni based superalloys, the constrained lattice misfit of alloys with high contents of Re and Ru even increases (more negative) with increasing temperature due to the strong dilution of the Re and Ru concentration in the γ phase [31]. However, at high temperatures, such as 1000 °C, in the 1$^{st}$-4$^{th}$ generation Ni based superalloys, their maximum lattice misfit modulus (less 0.4 % [31]) is lower than the one of our new alloys, CoNi-S3 and CoNiCr-2Ta. **Fig. 15 (a)** shows the rafted microstructure in alloy CoNiCr-2Ta after the creep test at 950 °C with an applied stress of 150 MPa after having accumulated ~3% strain during 1000 h. In **Fig. 15 (b)**, the less dislocations are able to cut through the rafts which is different from alloys Co9Al9W-0.12B [52] and alloy Co-9Al-7.5W-2Ta [28], in which dislocations can easily cut precipitates and leave a high number of planar defects, such as superlattice stacking faults and APBs. In addition, a typical dense dislocation network formed along the interface between γ' and γ, which is promoted as it relieves the high misfit stress. This indicates that the γ' phase is not easily cut and therefore dislocations have to climb and glide along the interfaces between the γ' and γ phase to circumvent the rafts. The inter-dislocation distance in the dislocation network will have a significant effect on creep life [49, 53]. If the γ'/γ lattice misfit is higher, the inter-dislocation distance is smaller, which is usually beneficial for creep strength. Accordingly, the new alloys should have an advantage in this respect relative to Ni based superalloys. However, in **Fig. 14(d)**, Ni based superalloys have proven their excellent high temperature capability, *i.e.*, super long creep life, especially at 1000 and 1050 °C, but the Ni based superalloys shown there were predominantly tested as single crystals. Therefore, in the future it is necessary to assess the new alloys' capability at high temperatures by employing single crystal.

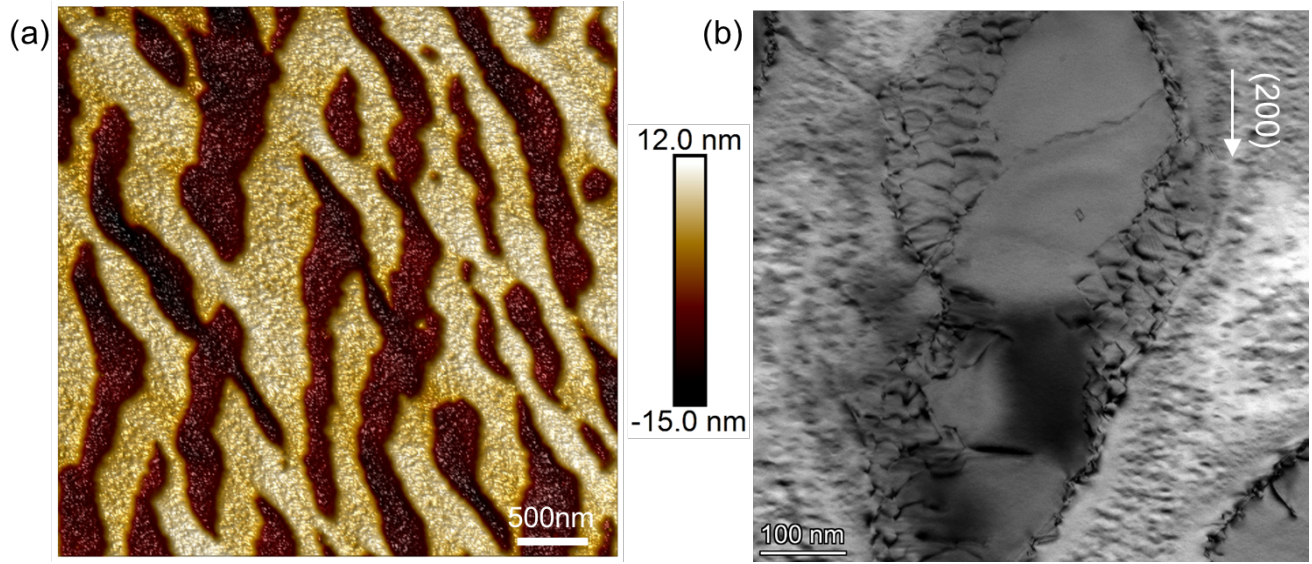

**Fig. 15.** (a) Rafted microstructure imaged by AFM in alloy CoNiCr-2Ta after creep test at 950 °C with an applied stress of 150 MPa (The bright gold is the γ′ phase and the dark red is the γ phase). (b) Typical dislocation network along interface between the γ' and γ phase in alloy CoNiCr-2Ta.

**4. Summary and outlook**

In this study, we investigated the effect of the Ti/Al ratio on the lattice misfit, yield stress, creep resistance and γ' phase stability, substituting Al by Ti in alloys Co-30Ni-(12.5-x)Al-xTi-2.5Mo-2.5W (at.%). With increasing Ti/Al ratio the γ'/γ lattice misfit increases and if the Ti/Al ratio ≥ 1, the lattice misfit becomes extremely high (≥ 1.3 %) but the γ' phase is still regular cubic which is different from a binary Co-12Ti alloy. In addition, high Ti/Al ratio improves the mechanical properties, such as higher yield stress and better creep resistance. With increasing Ti content, the γ' phase coarsening rate constant becomes smaller during aging at 900 °C due to reduced interfacial energy between phases γ and γ′ and higher lattice misfit. However, at a temperature of 1100 °C, the cubic γ' precipitates decomposed into deleterious η phases with D024 structure if the Ti content is too high, *e.g.*, Ti/Al ratio > 1.

Therefore, a new alloys design was derived, which employed an appropriate Ti/Al ratio, *i.e.*, Ti/Al ratio ≤ 1, and a high content Cr and Ta. Due to the high Ti/Al ratio in alloys, they exhibited an extremely high lattice-misfit with regular cubic $L1_2$ particles and excellent creep resistance. For example, the newly designed alloy CoNi-S3 has a higher γ′ solvus temperature (> 1200 °C) and an equally high lattice misfit (> 1.2 %) as the binary Co-12Ti superalloy but more regular cubic γ′ precipitates and better creep resistance than superalloys Co-9Al-9W and Co-9Al-9W-2Ti at 850 and 950 °C. In addition, compared with Ni based superalloys, the new alloys CoNiCr-2Ta and CoNi-S3 have higher lattice misfit at high temperatures, *e.g.*, 1000 and 1100 °C. However, single crystal Ni based superalloys have proven to show excellent creep resistance at high temperatures above 1050 °C.

Accordingly, the capability of the newly designed alloys at high temperatures, *e.g.*, 1000 and 1050 °C, still needs to be further explored by using single crystal and add additional refractory elements to strengthen the γ matrix. In addition, in our new alloys, we only considered high lattice misfit and limited additions of W and Ta. The competing single crystal Ni-base superalloys usually are much more complex multicomponent materials. A further improvement of the alloys presented in the current work could be achieved by addition of refractory elements, such as Re, Mo and Ru. These refractory elements not only have low diffusion coefficients in Ni and Co but can also significantly strengthen the γ matrix, which should have adorable effect on high temperature creep resistance.

## 5. Acknowledgement

We thank Dr. Jonathan Paul for his help with creep and compression tests and Dr. Ragle Raudsepp for atomic force microscopy characterisations, both at Helmholtz-Zentrum Hereon.

# Extreme high lattice-misfit superalloys with regular cubic L1$_2$ particles and excellent creep resistance

Zhida Liang[1,2,*], Andreas Stark[1], Florian Pyczak[1]

1. Institute of Materials Physics, Helmholtz-Zentrum Hereon, Max-Planck-Strasse 1, Geesthacht 21502, Germany
2. Laboratory for Electron Microscopy, Karlsruhe Institute of Technology, Engesserstraße 7, Karlsruhe 76131, Germany

∗ Corresponding author: Zhida Liang, zhida.liang@outlook.com

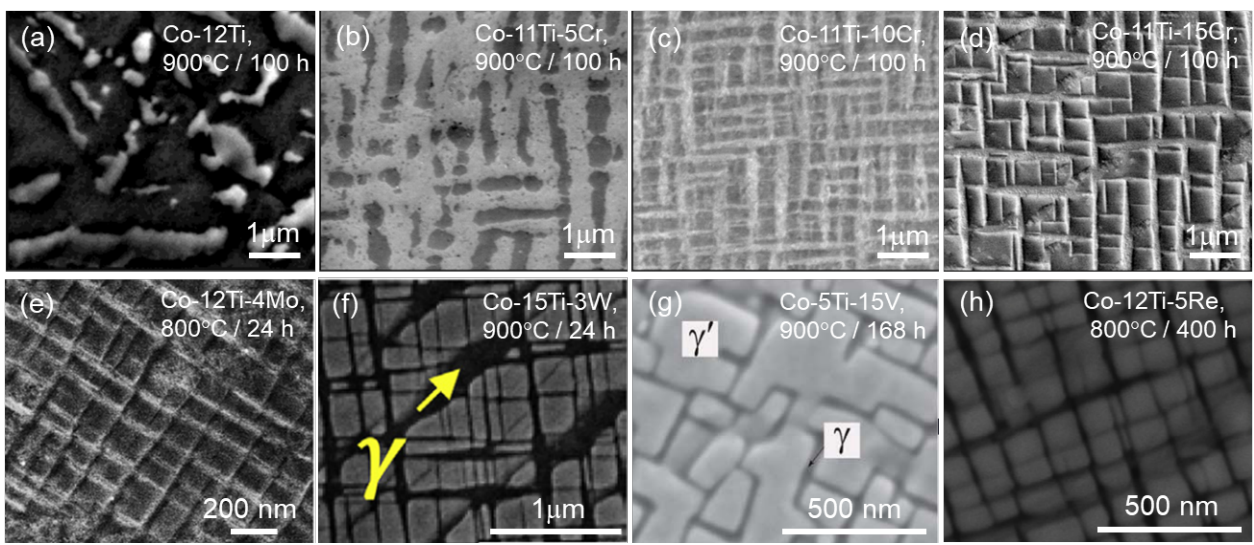

**Fig. S1**. Microstructure of the γ/γ′ Co-Ti-based superalloys: (a) Co-Ti system [1]; (b)-(d) Co-Ti-Cr system [2]; (e) Co-Ti-Mo system [3]; (f) Co-Ti-W system [4]; (g) Co-Ti-V system [5]; (h) Co-Ti-Re system [6].

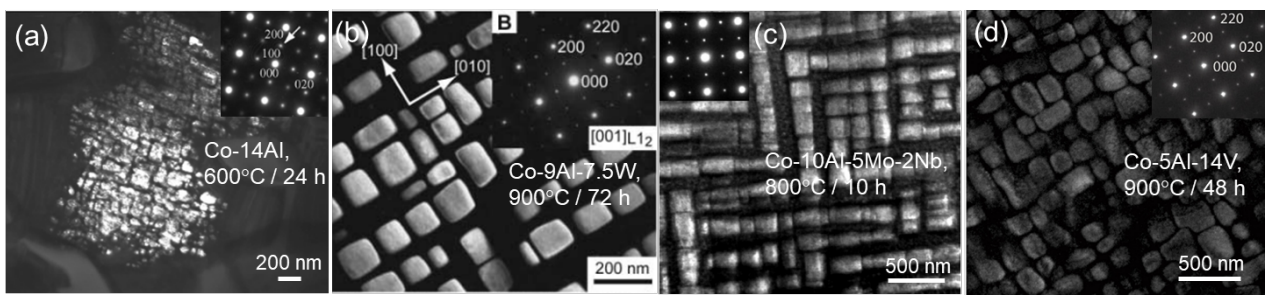

**Fig. S2.** Microstructure of the γ/γ′ Co-Al-based superalloys: (a) Co-Al system [7]; (b) Co-Al-W system [8]; (c) Co-Al-Mo system [9]; (d) Co-Al-V system [10].

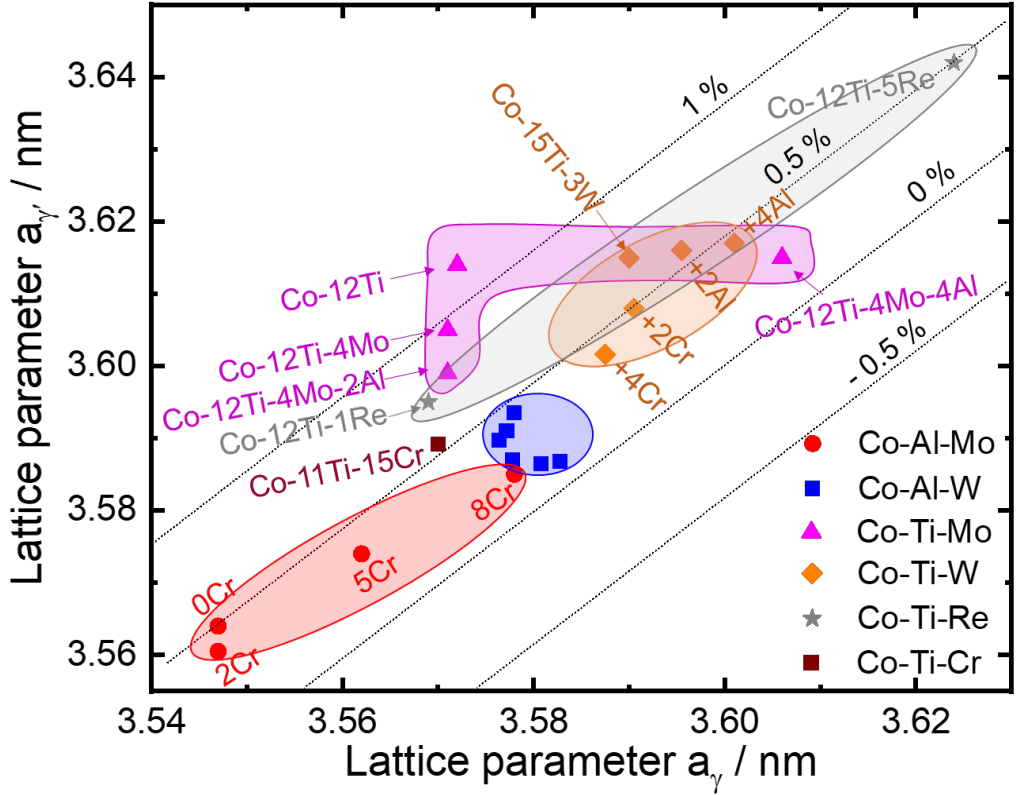

**Fig. S3.** Lattice misfit of γ/γ′ Co-Al-based and Co-Ti-based superalloys: Co-Al-W system [8]; Co-Al-Mo system [9]; Co-Ti-Mo system [3]; Co-Ti-W system [4]; Co-Ti-Re system [6]; Co-Ti-Cr system [2].

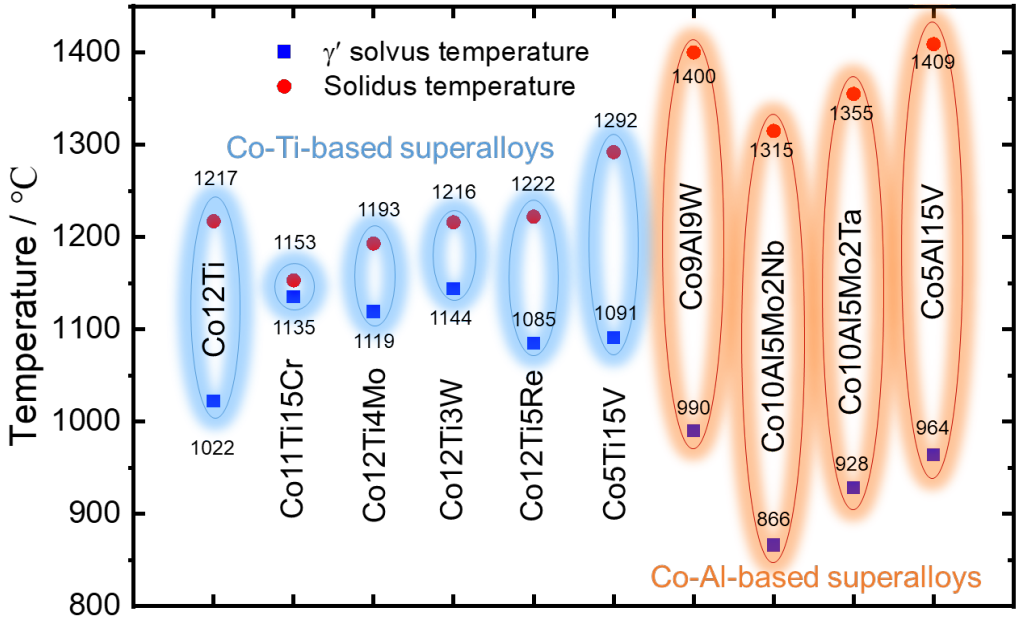

**Fig. S4.** Solution heat treatment temperature window ($T_{window} = T_{solidus} - T_{γ′\ solvus}$) of Co-Ti-based superalloys [1-6] and Co-Al-based superalloys [7-10].

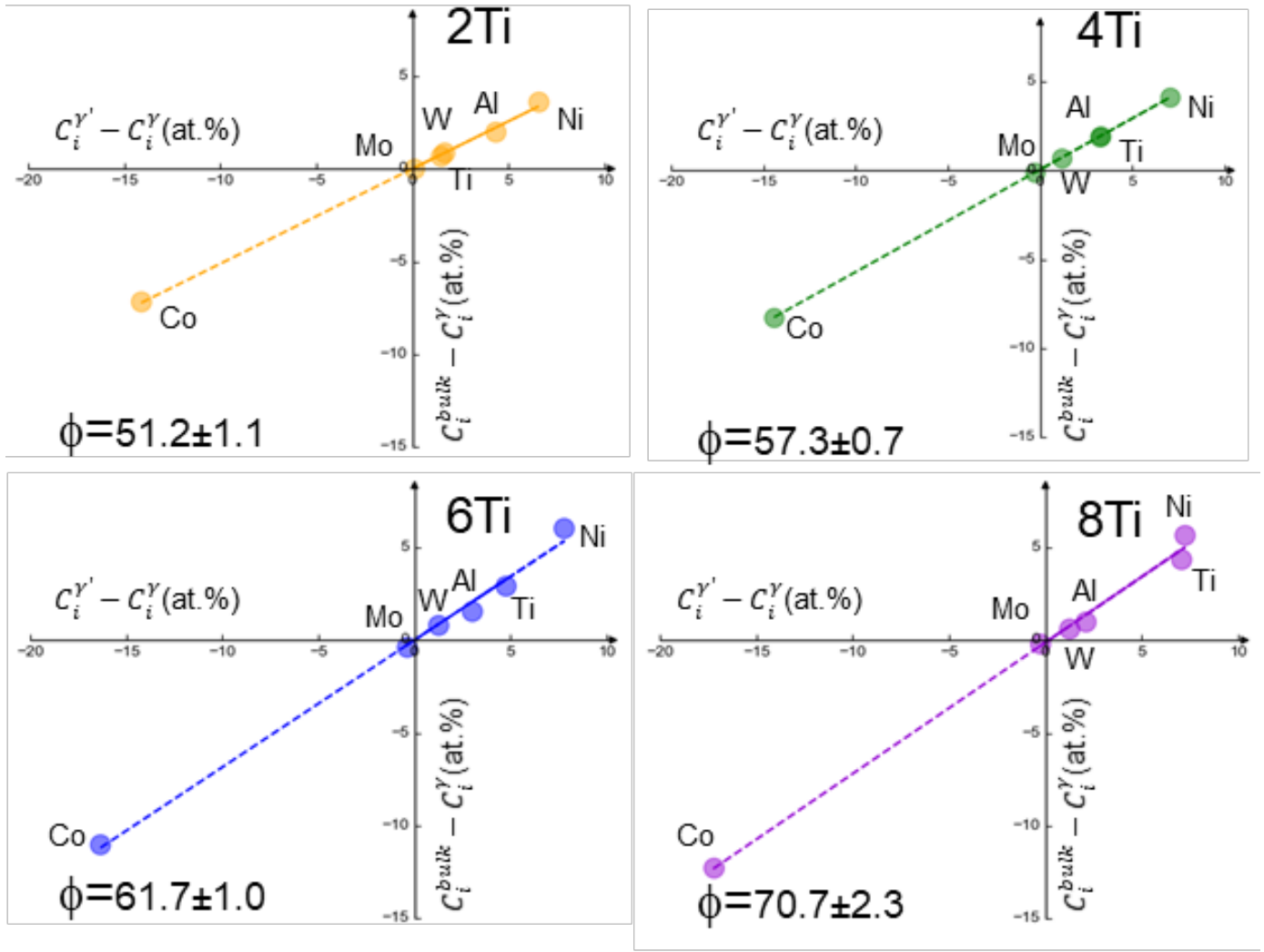

**Fig. S5.** Scatter plot with linear fitting between $(C_i^{\gamma'} - C_i^{\gamma})$ and $(C_i^{bulk} - C_i^{\gamma})$ for 2Ti, 4Ti, 6Ti and 8Ti alloys for estimation of γ′ volume fraction.

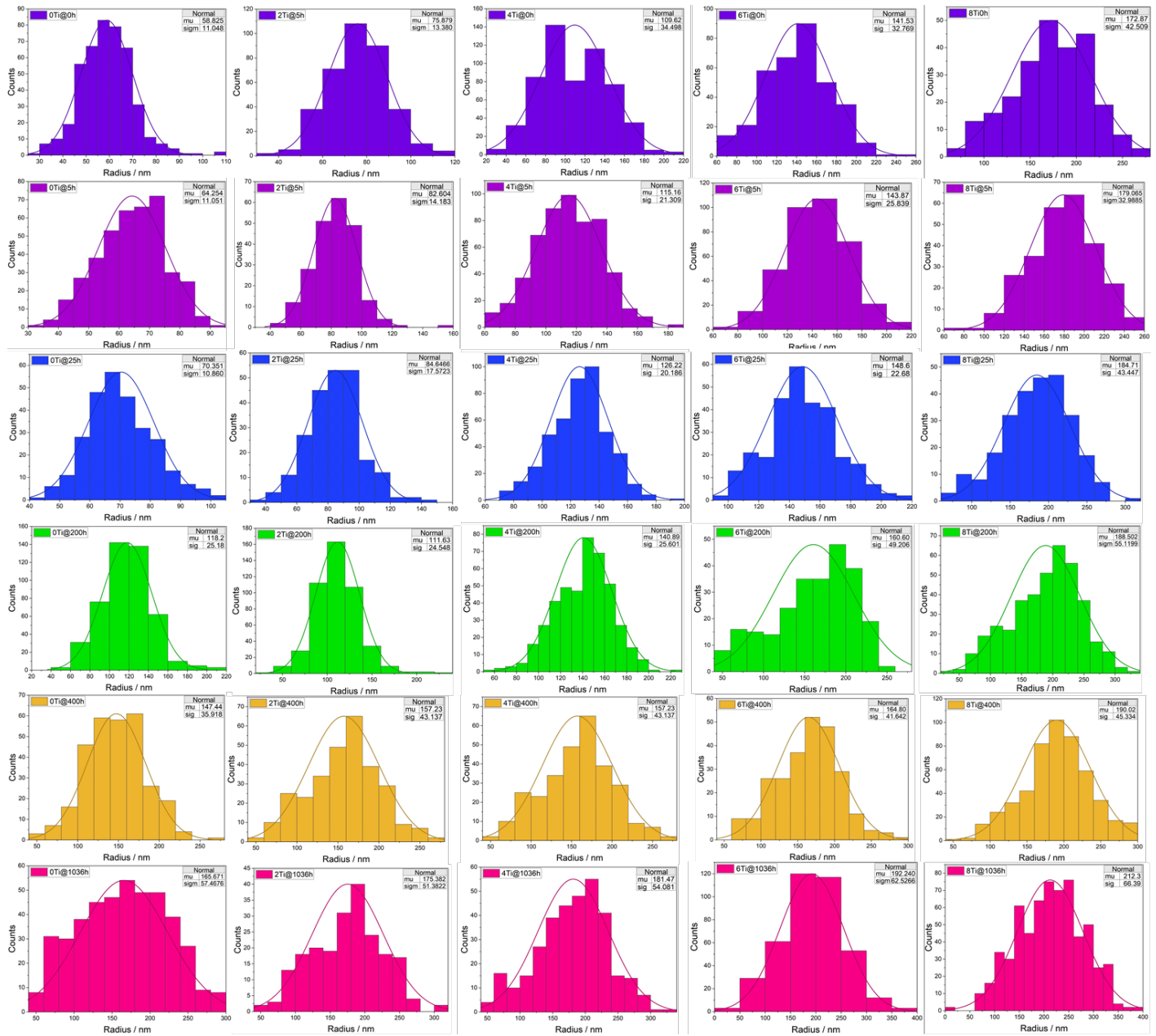

**Fig. S6.** Particle size distributions and their gaussian fittings of alloys 0Ti, 2Ti, 4Ti, 6Ti, and 8Ti during aging at 900 °C for 0, 5, 25, 200, 400, and 1036 h.